\documentclass[journal=jctcce,manuscript=article]{achemso} 
\usepackage[version=3]{mhchem} 
\usepackage{graphicx}
\usepackage{amssymb}
\usepackage{xcolor}

\newcommand{\pvnt}{P_v(\tilde{N})}

\newcommand{\fvnt}{F_v(\tilde{N})}

\newcommand{\Nt}{\tilde{N_v}}
\newcommand{\avgNt}{\langle \tilde{N_v} \rangle}
\newcommand{\varNt}{\langle \delta \tilde N_v^2 \rangle}

\newcommand{\kbt}{k_{\rm{B}} T}
\newcommand{\Rbar}{\overline{\textbf{R}}}

\title{Sparse Sampling of Water Density Fluctuations near Liquid-Vapor Coexistence}
\author{Erte Xi}
\author{Sean M. Marks}
\author{Suruchi Fialoke}
\author{Amish J. Patel}
\affiliation[University of Pennsylvania]
{Department of Chemical and Biomolecular Engineering, University of Pennsylvania, Philadelphia, PA 19104, USA}
\email{amish.patel@seas.upenn.edu}

\begin{document}

\begin{abstract}

\setlength{\baselineskip}{18pt}

The free energetics of water density fluctuations in bulk water, at interfaces, and in hydrophobic confinement inform the hydration of hydrophobic solutes as well as their interactions and assembly.
The characterization of such free energetics is typically performed using enhanced sampling techniques such as umbrella sampling.
In umbrella sampling, order parameter distributions obtained from adjacent biased simulations must overlap in order to estimate free energy differences between biased ensembles.
Many biased simulations are typically required to ensure such overlap, which exacts a steep computational cost.
We recently introduced a sparse sampling method, 
which circumvents the overlap requirement by using thermodynamic integration 
to estimate free energy differences between biased ensembles.
Here we build upon and generalize sparse sampling for characterizing the free energetics of 
water density fluctuations in systems near liquid-vapor coexistence.
We also introduce sensible heuristics for choosing the biasing potential parameters and strategies for adaptively refining them, 
which facilitate the estimation of such free energetics accurately and efficiently.
We illustrate the method by characterizing the free energetics of cavitation in 
a large volume in bulk water.
We also use sparse sampling to characterize the free energetics of capillary evaporation for water confined between two hydrophobic plates.
In both cases, sparse sampling is nearly two orders of magnitude faster than umbrella sampling.
Given its efficiency, the sparse sampling method is particularly well suited for characterizing 
free energy landscapes for systems wherein umbrella sampling is prohibitively expensive.
\end{abstract}\
Keywords: free energy method, umbrella sampling, thermodynamic integration, hydrophobic confinement

\setlength{\baselineskip}{18pt}

\section{Introduction}
%
At ambient conditions, water and many other liquids are close to coexistence with their vapor phase~\cite{stillinger,LCW,Chandler:Nature:2005,ashbaugh_SPT,LLCW,pgd:JPCL:2011,Suri:PRL:2014,Suri:PNAS:2016,Xi:PNAS:2016,desgranges2016free}.
Liquid water in the vicinity of hydrophobic surfaces is destabilized further, 
situating interfacial waters at the edge of a dewetting transition, 
and rendering them sensitive to unfavorable perturbations~\cite{rossky84,pgd:JPCB:2007,mittal_pnas08,Godawat:PNAS:2009,Giovambattista:PNAS:2009,Patel:JPCB:2010,Rotenberg:JACS:2011,Patel:PNAS:2011,Patel:JPCB:2012}.
Moreover, when confined between hydrophobic surfaces, liquid water can become metastable (or even unstable) with respect to its vapor~\cite{Luzar:JCP:2000,hummer_nanotube,Luzar:PRL:2003,Patankar:Langmuir:2004,bolhuis,pgd06,Giovambattista:JPCC:2007,chou_dewet,rasaiah2008water,Berne:JPCC:2009,berne_rev09,Mittal:Faraday:2010,Wang:PNAS:2011,Giacomello:PRL:2012,Altabet:JCP:2014,Remsing:PNAS:2015,Prakash:PNAS:2016,Altabet:PNAS:2017}.
Such interfacial or confined waters, which are situated near liquid-vapor coexistence,
play an important role in diverse processes, ranging from colloidal assembly and protein interactions, 
to heterogeneous vapor nucleation and the loss or recovery of superhydrophobicity~\cite{chaimovich2013length,wang2017colloidal,berne04,Freed:JPCB:2011,Snyder:PNAS:2011,Wang-Berne:PNAS:2011,Prakash:PNAS:2016,Patankar:Langmuir:2004,Shahraz:Langmuir:2012,Checco:PRL:2014,savoy2012molecular}.
In particular, displacing the interfacial or confined waters can incur a substantial free energetic cost,
and influence the kinetics of these process~\cite{Sharma:PNAS:2012,Sharma:JPCB:2012,Remsing:PNAS:2015,desgranges2017free}.
Thus, characterizing the free energetics of water density fluctuations  
can shed light on the mechanistic pathways associated with wetting-dewetting transitions in such systems,
and also inform the magnitude of the associated barriers~\cite{ChandlerPolymerLattice,Miller:PNAS:2007,Setny:PNAS:2013,weiss2017principles,Giacomello:PRL:2012,Checco:PRL:2014,Giacomello:JCP:2015,menzl2016molecular,Prakash:PNAS:2016}.

To characterize the free energetics of rare water density fluctuations, 
it becomes necessary to employ enhanced sampling techniques, 
such as umbrella sampling,
which are computationally expensive.
However, umbrella sampling can become prohibitively expensive when the simulations themselves are very expensive or when a large number of biased simulations are needed~\cite{zanjani2016colloidal,wang2017colloidal,kundu2011application,ray2017importance,hassanali2013proton,remsing2017dependence,sosso2016role}.
To address these challenges and facilitate the computationally efficient characterization of the free energetics of water density fluctuations, 
here we build upon a sparse sampling method that we previously introduced~\cite{Xi:JCTC:2016}, and generalize it to study systems near liquid-vapor coexistence.
%
%
In the following sections, we will first use model (analytical) free energy landscapes to illustrate the
sparse sampling method, the reasons underlying its successes, and situations when it is challenged.
We then extend the sparse sampling method to tackle such challenging scenarios.  
Both the accuracy and the efficiency of sparse sampling rely on the choice of biasing potential parameters;
we introduce sensible heuristics for choosing such parameters and strategies for adaptively refining them.
We then highlight the efficiency of the sparse sampling method by using it to characterize the free energetics of water density fluctuations in a large volume in bulk water. 
Finally, we apply sparse sampling to a particularly challenging system, 
wherein water is confined between two hydrophobic surfaces, 
and the free energetics of water density fluctuations display 
two basins that are separated by a barrier.

\section{Illustrating Sparse Sampling Using Model Landscapes}

\subsection{Umbrella Sampling {\it vs} Sparse Sampling}
%
Characterizing the free energetics of order parameter fluctuations, which are too rare to be observed in unbiased molecular simulations, requires the use of non-Boltzmann sampling techniques such as umbrella sampling.
To facilitate the sampling of such rare order parameter fluctuations, 
umbrella sampling prescribes the use of biasing potentials, 
which enhance the likelihood with which otherwise improbable fluctuations are sampled.
Although umbrella sampling provides a powerful way to characterize the free energetics of rare order parameter fluctuations, 
it exacts a steep computational cost, which can become prohibitive,
either when the biased simulations themselves are expensive~\cite{zanjani2016colloidal,wang2017colloidal,kundu2011application,ray2017importance,remsing2014role,sosso2016role,remsing2017dependence}, or when a large number of biased simulations are needed to span the order parameter range of interest.
To address this challenge, we recently introduced a sparse sampling method~\cite{Xi:JCTC:2016}, 
which employs sparsely separated biased simulations,
and can be orders of magnitude more efficient than conventional umbrella sampling. 

Although the sparse sampling method is generally applicable to any order parameter, 
here we focus on an order parameter that represents the smoothed (or coarse-grained) 
number of water molecules, $\Nt$, in an observation volume, $v$, of interest.
We bias $\Nt$ in lieu of the closely related (discrete) number of waters, $N_v$, in $v$,
because performing biased molecular dynamics (MD) simulations 
with a discrete order parameter results in impulsive forces.
The precise definition of $\Nt$ as well as its dependence on the atomic positions, $\Rbar$, can be found in ref.~\cite{Patel:JSP:2011}. 
We wish to estimate the free energetics of water density (or rather $\Nt$) fluctuations, $\fvnt = -\kbt \ln \pvnt$, where 
$k_{\rm B}$ is the Boltzmann constant, $T$ is the system temperature, and
$\pvnt = \langle \delta(\Nt - \tilde N) \rangle_0$, is the probability of observing $\tilde N$ coarse-grained waters in $v$;
here, $\langle \mathcal{O}(\Rbar) \rangle_0 \equiv \int d\Rbar~\mathcal{O}(\Rbar) \exp(-\beta \mathcal{H}_0) / Q_0$ represents the ensemble average of $\mathcal{O}(\Rbar)$, 
given the generalized Hamiltonian, $\mathcal{H}_0(\Rbar)$,
and $Q_0\equiv \int d\Rbar~\exp(-\beta \mathcal{H}_0)$ is the corresponding partition function.
To facilitate the sampling of $\Nt(\Rbar)$ over the entire $\tilde N$-range of interest, we use potentials, $U_{\bar\lambda} (\Nt)$, 
which bias $\Nt$, and are parametrized by the vector, $\bar\lambda$;
that is, we perform biased simulations using the Hamiltonians, $\mathcal{H}_{\bar\lambda}(\Rbar) =  \mathcal{H}_{0}(\Rbar) + U_{\bar\lambda} (\Nt(\Rbar))$.
Using such biased simulations, we can estimate averages in the biased ensembles, $\langle \mathcal{O}(\Rbar) \rangle_{\bar\lambda}$, and in particular, 
we estimate the biased distributions, $P_v^{\bar\lambda}(\tilde N) = \langle \delta(\Nt - \tilde N) \rangle_{\bar\lambda}$, 
and the corresponding free energetics, $F_v^{\bar\lambda} (\tilde N) = -\kbt \ln P_v^{\bar\lambda}(\tilde N)$.
For the $\tilde N$-range sampled by a biased simulation, $\fvnt$ can then be related to $F_v^{\bar\lambda} (\tilde N)$ using the exact result 
\begin{equation}
\fvnt =  F_v^{\bar\lambda} (\tilde N) - U_{\bar\lambda} (\tilde N) + F_{\bar\lambda},
\label{eq:start-lambda}
\end{equation}
where $F_{\bar\lambda} \equiv -\kbt \ln \left(\frac{Q_{\bar\lambda}}{Q_0}\right)$ is the free energy difference between the biased and the unbiased ensembles, and $Q_{\bar\lambda}$ is the partition function corresponding to $\mathcal{H}_{\bar\lambda}$. 
The derivation of Equation~\ref{eq:start-lambda} is included as supplementary information.
Using $F_v^{\bar\lambda} (\tilde N)$ obtained from the biased simulation, and the known functional form of $U_{\bar\lambda} (\tilde N)$, 
$\fvnt$ can then be obtained to within the unknown constant offset, $F_{\bar\lambda}$.
%

In umbrella sampling, estimates of $F_{\bar\lambda}$ for the different biased ensembles are obtained by 
requiring an overlap in the range of $\tilde N$-values sampled by adjacent biased simulations,
and in particular, by matching $\fvnt$-values obtained from different biased simulations in the overlap regions; 
algorithms such as the Weighted Histogram Analysis Method (WHAM) or Multi-state Bennet Acceptance Ratio (MBAR)
are typically used for this purpose~\cite{Kumar:JCC:1992, Roux:CPC:2001,MBAR,zhu2012convergence,UWHAM,stelzl2017dynamic}.
Such an umbrella sampling strategy is powerful and has facilitated the characterization of the free energetics of water density fluctuations in numerous contexts~\cite{Patel:JPCB:2010,Rotenberg:JACS:2011,Patel:PNAS:2011,LLCW,Patel:JPCB:2012,Remsing:JCP:2015}.
However, to satisfy the above overlap requirement, it becomes necessary to run many long biased simulations.
In practice, the associated computational cost limits the size of $v$ as well as the level of simulation detail (e.g., classical vs ab initio MD~\cite{sosso2016role,remsing2017dependence}) for which $\fvnt$ can be characterized.
In contrast, the sparse sampling method does not require overlap between adjacent biased distributions~\cite{Xi:JCTC:2016}.
Instead, sparse sampling employs thermodynamic integration to obtain $F_{\bar\lambda}$,
and in principle, it can facilitate estimation of $\fvnt$ at sparsely distributed $\tilde N$-values orders of magnitude faster than conventional umbrella sampling.
Below, we first describe the sparse sampling method, as introduced in ref.~\cite{Xi:JCTC:2016}, and discuss its advantages and shortcomings; then, we generalize the sparse sampling method to address those shortcomings.

\subsection{Sparse Sampling with a Linear Biasing Potential}
%
In ref.~\cite{Xi:JCTC:2016}, we introduced sparse sampling using a linear biasing potential, $U_\phi (\Nt) = \phi \Nt$, parameterized by $\phi$; that is, $\bar\lambda=[\phi]$, and the corresponding biased ensemble Hamiltonian, $\mathcal{H}_{\phi} = \mathcal{H}_0 + U_{\phi} (\Nt)$. 
Equation~\ref{eq:start-lambda} then becomes:
\begin{equation}
F_v(\tilde N) =  F_v^{\phi} (\tilde N) - U_{\phi} (\tilde N) + F_{\phi}.
\label{eq:start-phi}
\end{equation}
Instead of relying on an overlap in the $\tilde N$-values sampled by adjacent biased ensembles, sparse sampling prescribes estimating $F_{\phi}$ by integrating $dF_{\phi} / d\phi \equiv \langle dU_\phi/d\phi \rangle_\phi = \avgNt_{\phi}$,
that is, by using the identity:
\begin{equation}
F_{\phi} = \int_0^{\phi} \avgNt_{\phi^{\prime}} d\phi^{\prime}. 
\label{eq:TI1}
\end{equation}
In practice, only a small number of biased simulations that sample well-separated $\Nt$-values are performed;
hence the term {\it sparse} sampling.
Following Equation~\ref{eq:TI1}, the corresponding $F_{\phi}$-estimates are then obtained by numerically integrating
the averages, $\avgNt_{\phi}$, obtained from those simulations.
To minimize integration errors in estimating $F_\phi$, the sparse sampling method thus relies on capturing the variation of $\avgNt_{\phi}$, 
the average thermodynamic force (and the integrand in thermodynamic integration), with $\phi$.
In other words, to accurately estimate $F_{\phi}$, the biased simulations must capture the functional dependence of $\avgNt_{\phi}$ on $\phi$.
Indeed, our choice of a linear biasing potential, $U_\phi (\Nt) = \phi \Nt$, was informed by this requirement.
In particular, because $d\avgNt_\phi / d\phi = - \beta \varNt_\phi \le 0$, this choice ensures that the average thermodynamic force, $\avgNt_\phi$, decreases monotonically with $\phi$, and for certain systems, the decrease can even be linear in $\phi$~\cite{Patel:JPCB:2014}. 
Here, $\varNt_\phi$ is the variance of $P_v^\phi(\tilde N)$.
In ref.\cite{Xi:JCTC:2016}, we used Equations~\ref{eq:start-phi} and~\ref{eq:TI1} to efficiently estimate $\fvnt$ in small volumes in bulk water and at interfaces, and in large volumes in heterogeneous interfacial environments such as protein hydration shells.
However, as highlighted in ref.~\cite{Xi:JCTC:2016}, and discussed in further detail below, the use of a linear potential is not suitable for sparse sampling systems near liquid-vapor coexistence.

\subsection{Sparse Sampling Systems Near and Far from Coexistence}
\begin{figure}[htbp]
\begin{center}
\vspace{-0.5in}
\includegraphics[width=0.99\textwidth]{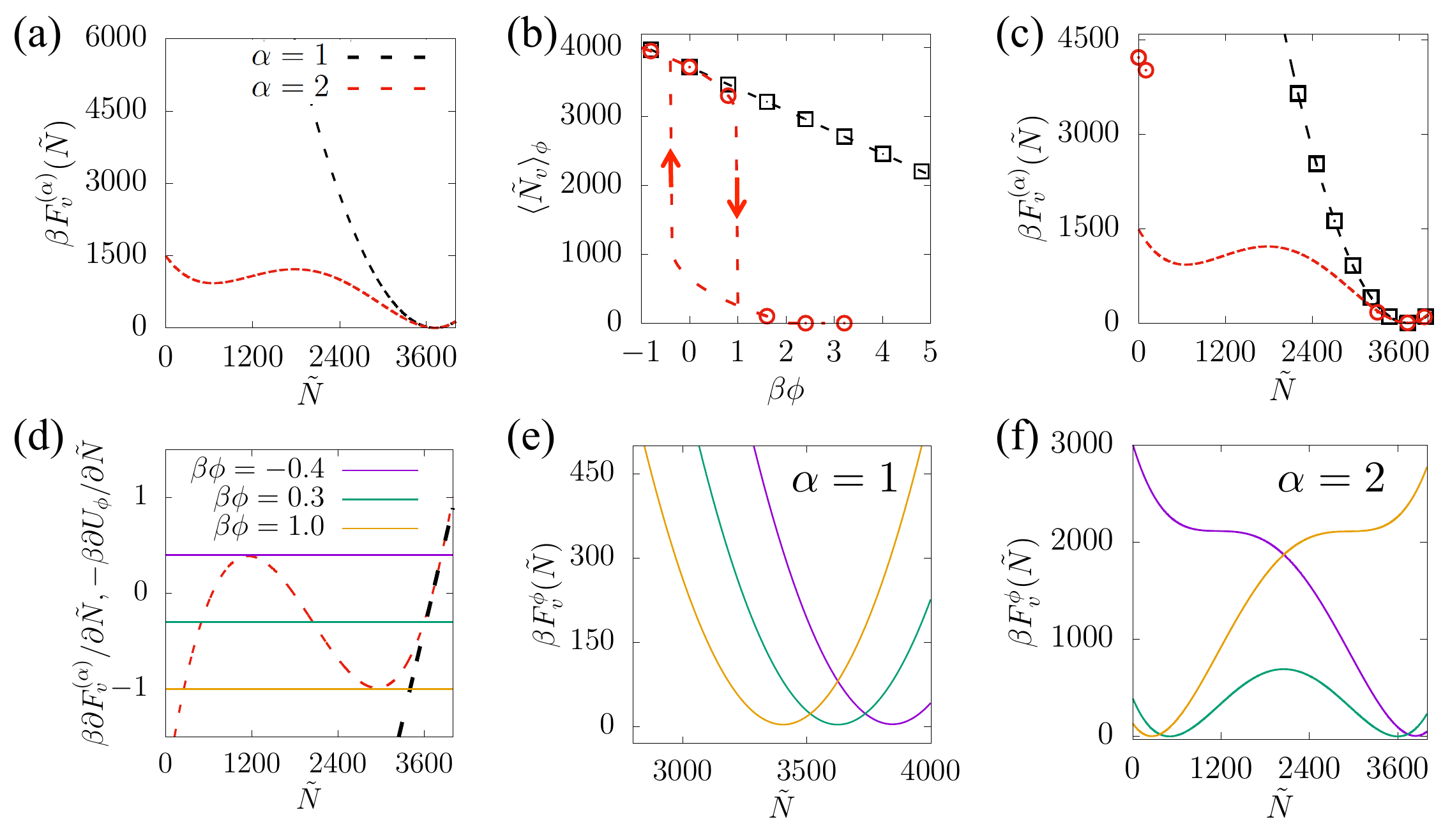}
\caption{
Sparse sampling with a linear biasing potential, $U_\phi(\Nt)=\phi\Nt$.
(a) Two model systems with distinctive free energy profiles are considered: system 1 displays a monostable free energy profile, $F_v^{\rm (1)}(\tilde N)$ (black), 
whereas system 2 is bistable, $F_v^{\rm (2)}(\tilde N)$ (red). 
Both systems have the same minimum and curvature in the liquid basin (high $\tilde N$), 
but $F_v^{\rm (2)}(\tilde N)$ deviates substantially from $F_v^{\rm (1)}(\tilde N)$ at lower $\tilde N$. 
(b) The response of the average number of waters, $\avgNt_\phi$, to the strength of the biasing potential, $\phi$, 
at evenly- (and sparsely-) spaced $\phi$-values is shown.
For system 1, $\avgNt_\phi$ decreases linearly with $\phi$, 
whereas $\avgNt_\phi$ displays a sharp decrease for system 2.
Moreover, $\avgNt_\phi$ displays hysteresis for system 2;
that is, for certain values of $\phi$, $\avgNt_\phi$ depends on 
whether the system was initialized in the dry state (upward arrow) or the wet state (downward arrow). 
(c) For system 1, the sparse sampled $\fvnt$-values (at $\tilde N = \avgNt_\phi$, black squares) accurately capture $F_v^{\rm (1)}(\tilde N)$ over its entire range.
In contrast, sparse sampling fails to characterize $\fvnt$ for system 2 (red circles) over a significant range of intermediate $\tilde N$-values. 
Moreover, the hysteresis in $\avgNt_\phi$ results in substantial errors in estimating $F_\phi$, 
which propagate to $\fvnt$ for low $\tilde N$-values.
(d) Extrema of the biased free energy profiles, $F_v^\phi(\tilde N)$, can be obtained from the intersection of $\partial F_v / \partial \tilde N$ and $-\partial U_\phi / \partial \tilde N = -\phi$.
For system 1, $\partial F_v / \partial \tilde N$ increases monotonically with $\tilde N$, and intersects $-\phi$ only once for all values of $\phi$, suggesting that the corresponding $F_v^\phi(\tilde N)$ are monostable.
For system 2, multiple (three) intersections are observed for certain $\phi$-values,
suggesting bistability in the corresponding $F_v^\phi(\tilde N)$.
(e) As expected, all the biased free energy profiles, $F_v^\phi(\tilde N)$, are monostable for system 1.
(f) In contrast, the biased free energy profiles for system 2 are bistable in the range, $-0.4 < \beta\phi < 1$,
and explain the hysteresis observed in panel b for this $\phi$-range.
}
\label{fig:caveat}
\end{center}
\end{figure}

To illustrate the differences between systems near and far from liquid-vapor coexistence,
we consider two model systems characterized by distinct analytical free energy profiles (Figure~\ref{fig:caveat}a).
System 1 is representative of a system far from liquid-vapor coexistence, and is monostable: 
$F_v^{\rm (1)}(\tilde N) = \frac{1}{2} \kappa_{\rm m} (\tilde N - n^{(1)}_{\rm liq})^2$; the corresponding distribution, $\pvnt$, is Gaussian.
In contrast, system 2 resembles a system close to liquid-vapor coexistence, and is bistable:
$F_v^{\rm (2)}(\tilde N) = \kappa_{\rm b} (\tilde N - n^{(2)}_{\rm vap})^2 (\tilde N - n^{(2)}_{\rm liq})^2 - \phi_{\rm b} \tilde N$, 
with distinct basins at low and high $\tilde N$-values.
We have designed $F_v^{\rm (1)}(\tilde N)$ and $F_v^{\rm (2)}(\tilde N)$ so that the high $\tilde N$ (liquid) basins for the two model systems
are similar to one another, and to a $3~$nm spherical observation volume in bulk water, which we will consider in the following section.
To do so, we choose 
$\beta\kappa_{\rm m} = 0.00318$, 
$n^{(1)}_{\rm liq}= 3720$, 
$\beta\kappa_{\rm b}=1.2 \times 10^{-10}$, 
$n^{(2)}_{\rm vap} = 500$,
$n^{(2)}_{\rm liq} = 3600$ and 
$\beta\phi_{\rm b}=0.3$,
where $\beta \equiv 1/\kbt$.
Although $F_v^{\rm (1)}(\tilde N)$ and $F_v^{\rm (2)}(\tilde N)$ are representative of commonly encountered free energy profiles 
far from and near liquid-vapor coexistence respectively, we note that these model systems are purely illustrative.
For typical systems of interest, such $\fvnt$-profiles will not be known a priori; rather, our goal will be to characterize them using sparse sampling.

The first step in characterizing the $\fvnt$-profiles of the above systems using sparse sampling is to obtain the averages, $\avgNt_\phi$, for several $\phi$-values. 
In lieu of performing biased simulations, we infer the biased free energetics, $F_v^\phi(\tilde N)$, 
by reweighting the $\fvnt$-profiles for the two systems using Equation~\ref{eq:start-phi}.
The locations of the minima in $F_v^\phi(\tilde N)$ then provide us with the averages, 
$\avgNt_\phi$, that would be obtained from the corresponding biased simulations.
We assume that all biased simulations are initialized with either high $\Nt$-values ($\Nt^0 \approx n^{(1)}_{\rm liq}$) or low $\Nt$-values ($\Nt^0 \approx 0$).
Thus, if $F_v^\phi(\tilde N)$ displays multiple minima separated by a barrier, 
we assign to $\avgNt_\phi$ the minimum that is closest to $\Nt^0$,
noting that in realistic simulations, 
a barrier of only a few $\kbt$ 
is sufficient to localize the system 
in the basin it was initialized in.
In this instance, $\avgNt_\phi$ does not represent a true biased ensemble average, but an estimate of the average that would be obtained from a biased simulation, and one that may suffer from serious inaccuracies due to the loss of ergodicity, which accompanies a bistable $F_v^\phi(\tilde N)$.

In Figure~\ref{fig:caveat}b (symbols), we show estimates of $\avgNt_\phi$, computed accordingly for equally spaced $\phi$-values.
Although $\avgNt_\phi$ decreases monotonically with $\phi$ for both systems, 
there are significant differences in the $\avgNt_\phi$ {\it vs.} $\phi$ response of the two systems.
For system 1, the dependence of $\avgNt_\phi$ on $\phi$ is particularly simple; $\avgNt_\phi$ decreases linearly with $\phi$.
Moreover, identical estimates of $\avgNt_\phi$ are obtained regardless of how the biased simulations are initialized.
The integral, $F_{\phi} = \int_0^{\phi} \avgNt_{\phi^{\prime}} d\phi^{\prime}$ (Equation~\ref{eq:TI1}), can thus be estimated accurately
for each of the 8 equally-spaced $\phi$-values.
Then, Equation~\ref{eq:start-phi} can be used to obtain estimates of $\fvnt$ at $\tilde N = \avgNt_\phi$.
These estimates are shown in Figure~\ref{fig:caveat}c, and are in excellent agreement with $F_v^{\rm (1)}(\tilde N)$. 
The linear dependence of $\avgNt_\phi$ on $\phi$ also means that 
the $\Nt$-values sampled in the biased simulations
are distributed across the $\tilde N$-range of interest,
enabling determination of $\fvnt$ over that entire range.
Thus, the sparse sampling method in conjunction with a linear potential is particularly well suited for characterizing the free energetics of systems far from coexistence.

In contrast with the gradual decrease of $\avgNt_\phi$ with increasing $\phi$ for system 1, $\avgNt_\phi$ decreases sharply over a narrow range of $\phi$-values for system 2, as shown in Figure~\ref{fig:caveat}b.
Thus, for uniformly spaced $\phi$-values, high and low $\Nt$-values are sampled well in the biased simulations, but intermediate values are not.
Consequently, there is a large range of $\tilde N$-values ($200 \lesssim \tilde N \lesssim 3400$) for which estimates of $\fvnt$ cannot be obtained (Figure~\ref{fig:caveat}c).
Moreover, the gap in our knowledge of the functional form of $\avgNt_\phi$ versus $\phi$ results in significant errors in our estimation of $F_\phi$ for the higher $\phi$-values, and thereby in estimates of $\fvnt$ for the lower $\tilde N$-values (Figure~\ref{fig:caveat}c).
To make matters worse, for the $\phi$-range over which $\avgNt_\phi$ decreases from high to low values, 
biased simulations initialized with low and high $\Nt$-values result in very different estimates of $\avgNt_\phi$; 
that is, hysteresis is observed in $\avgNt_\phi$ versus $\phi$ (Figure~\ref{fig:caveat}b).

%
To understand this contrast between systems 1 and 2, we recognize that the
$\Nt$-values sampled in a biased simulation will be in the vicinity of a minimum in $F^{\phi}_v(\tilde N)$.
Extrema of $F^{\phi}_v(\tilde N)$ obey $\partial F^{\phi}_v / \partial \tilde N = 0$, 
and because $F^{\phi}_v(\tilde N) = F_v(\tilde N) + U_\phi(\tilde N) - F_\phi$ (Equation~\ref{eq:start-phi}),
the extrema will satisfy $\partial F_v / \partial \tilde N = -\partial U_\phi / \partial \tilde N = -\phi$.
In Figure~\ref{fig:caveat}d, we plot $\partial F_v / \partial \tilde N$ for both systems; for system 1, it is linear and increases monotonically with $\tilde N$, whereas for system 2, it varies non-monotonically with $\tilde N$. 
For system 1, there can thus be only one solution to $\partial F_v / \partial \tilde N = -\phi$, and thereby only one minimum in $F^{\phi}_v(\tilde N)$ for any value of $\phi$ (Figure~\ref{fig:caveat}d).
This is indeed the case as shown in Figure~\ref{fig:caveat}e.
The corresponding biased distributions, $P^{\phi}_v(\tilde N)$, will thus be unimodal, facilitating the efficient estimation of the thermodynamic force, $\avgNt_\phi$, for all values of $\phi$.
Moreover, these arguments apply to any system for which $\partial F_v / \partial \tilde N$ increases monotonically with $\tilde N$, and correspondingly, $\fvnt$ is convex over the entire range of $\tilde N$-values.
In ref.~\cite{Xi:JCTC:2016}, we showed that heterogeneous surfaces, such as the protein hydration shells, which display a wide range of chemistries from hydrophobic to hydrophilic, indeed satisfy these criteria.

In contrast, the non-monotonic variation of $\partial F_v / \partial \tilde N$ with $\tilde N$ for system 2 
allows for the possibility of 
three solutions to $\partial F_v / \partial \tilde N = -\phi$  
in the range $\phi_{\rm min} \le \phi \le \phi_{\rm max}$ (with $\beta\phi_{\rm min}=-0.4$ and $\beta\phi_{\rm max}=1$).
In this $\phi$-range, $F^{\phi}_v(\tilde N)$ exhibits two minima and a maxima, that is, two basins separated by a barrier, as shown in Figure~\ref{fig:caveat}f.
The presence of a sufficiently large barrier in $F^{\phi}_v(\tilde N)$ will localize the system to the basin it was initialized in, leading to hysteresis and erroneous estimates of $\avgNt_\phi$.
Moreover, for the wide range of $\tilde N$-values for which $\partial F_v / \partial \tilde N$ decreases, $\partial^2 F^{\phi}_v / \partial \tilde N^2 = \partial^2 F_v / \partial \tilde N^2 < 0$.
As a result, $F^{\phi}_v(\tilde N)$ will have negative curvature in this $\tilde N$-range, 
precluding its sampling in the biased simulations, 
and resulting in a sharp decrease in $\avgNt_\phi$ as $\phi$ is increased from $\phi_{\rm min}$ to $\phi_{\rm max}$.   
These issues arise from the presence of locally concave regions of $\fvnt$, and
hinder the sparse sampling method from obtaining accurate estimates of $\fvnt$ over the entire $\tilde N$-range of interest. 
Because concave regions in $\fvnt$ are a common feature of systems near coexistence,
the use of a linear potential is not appropriate for sparse sampling such systems. 

\subsection{Extending Sparse Sampling to Systems near Coexistence}
\begin{figure}[htbp]
\begin{center}
\includegraphics[width=0.99\textwidth]{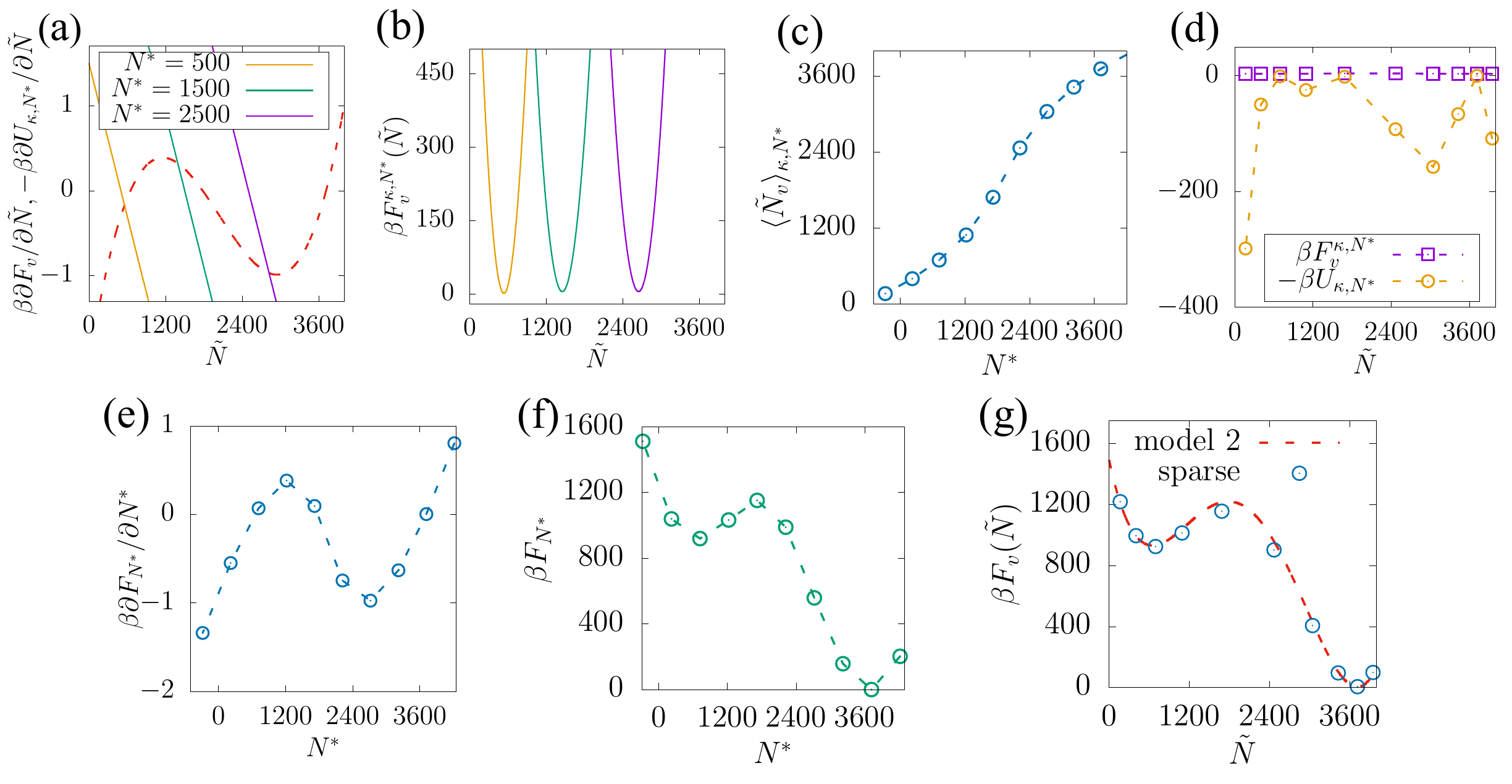}
\caption{
Sparse sampling system 2 using a harmonic potential, $U_{\kappa,N^*}(\Nt) = \frac{\kappa}{2}(\Nt - N^*)^2$. 
(a) Extrema of the biased free energy profiles, $F_v^{\kappa, N^*}(\tilde N)$, can be obtained from the intersection of $\partial F_v / \partial \tilde N$ and $-\partial U_{\kappa,N^*} / \partial \tilde N = -\kappa (\tilde N - N^*)$.
For sufficiently large $\kappa$, only one intersection will be observed for all $N^*$-values, 
suggesting that the biased free energetics ought to be monostable.
(b) The biased free energy profiles, $F_v^{\kappa, N^*}(\tilde N)$ are indeed monostable, as expected.
(c) The average number of waters, $\avgNt_{\kappa, N^*}$ in $v$, increases monotonically with $N^*$. 
(d) The first two terms on the right hand side of Equation~\ref{eq:start-kappa}, $F_v^{\kappa, N^*}(\tilde N)$ and $-U_{\kappa,N^*}(\tilde N)$, are shown for  
$\tilde N = \avgNt_{\kappa,N^*}$.
(e) The average thermodynamic force, which must be integrated to obtain $F_{\kappa, N^*}$, 
the free energy difference between biased and unbiased ensembles, 
is shown as a function of $N^*$.
(f) The free energy differences, $F_{N^*} \equiv F_{\kappa, N^*} - F_{\kappa, N^*_{\rm ref}}$, obtained by integrating the results in panel e, are shown for the $N^*$-values considered here.
(g) The sparse sampled $\fvnt$-values (at $\tilde N = \avgNt_{\kappa,N^*}$) accurately capture $F_v^{\rm (2)}(\tilde N)$ over its entire range.
}
\label{fig:harmonic}
\end{center}
\end{figure}

%
To generalize the sparse sampling method for systems near coexistence, here we explore the use of a biasing potential with a different (non-linear) functional form. 
In particular, we consider the frequently used parabolic potential, $U_{\kappa, N^*} (\Nt) = \frac{1}{2} \kappa (\Nt - N^*)^2$, 
which is parameterized by $\bar\lambda=[\kappa,N^*]$.
For sufficiently large $\kappa$-values, such a biasing potential can regularize the corresponding biased free energy profile, $F_v^{\kappa, N^*}(\tilde N)$, ensuring that it is monostable.
To illustrate this point, we first rewrite Equation~\ref{eq:start-lambda} for this potential:  
\begin{equation}
\fvnt = F_v^{\kappa, N^*}(\tilde N) - U_{\kappa, N^*} (\tilde N)  + F_{\kappa, N^*}
\label{eq:start-kappa}
\end{equation}
For $F_v^{\kappa, N^*}(\tilde N)$ to be monostable, $dF^{\kappa,N^*}_v / d\tilde N = 0$ ought to have only one solution, 
and correspondingly, so should $dF_v / d\tilde N = - dU_{\kappa, N^*} / d\tilde N = -\kappa(\tilde N - N^*)$.
In Figure~\ref{fig:harmonic}a, we again plot $dF^{(2)}_v / d\tilde N$ for system 2 (as in Figure~\ref{fig:caveat}d);
however, we now search for its intersection with $-\kappa(\tilde N - N^*)$ with $\beta \kappa = 0.003$.
In particular, we see that lines corresponding to $-\kappa(\tilde N - N^*)$ for several different $N^*$-values 
intersect $dF^{(2)}_v / d\tilde N$ only once when a sufficiently large $\kappa$-value is chosen.
As a result, the corresponding biased free energy profiles all display a single basin (Figure~\ref{fig:harmonic}b).
Moreover, the biased simulations facilitate sampling of $\Nt$ over the entire range of $\tilde N$-values, and the corresponding $\avgNt_{\kappa, N^*}$-values increase systematically with $N^*$ (Figure~\ref{fig:harmonic}c).
In fact, the monotonic increase of $\avgNt_{\kappa, N^*}$ with $N^*$ is guaranteed because $\partial \avgNt_{\kappa, N^*} / \partial N^* = \beta\kappa \varNt_{\kappa, N^*} \ge 0$.
%

Using Equation~\ref{eq:start-kappa}, we can then estimate $\fvnt$ at the $\avgNt_{\kappa, N^*}$-values shown in Figure~\ref{fig:harmonic}c. 
We can obtain $F_v^{\kappa, N^*}(\tilde N)$ directly from the biased simulations, 
and to a good approximation, $\beta F_v^{\kappa, N^*}(\tilde N = \avgNt_{\kappa, N^*}) \approx \frac{1}{2} \log(2\pi\varNt_{\kappa,N^*})$.
Similarly, $U_{\kappa, N^*}(\tilde N)$ is readily obtained as: $U_{\kappa, N^*}(\tilde N = \avgNt_{\kappa, N^*}) = \frac{1}{2}\kappa(\avgNt_{\kappa, N^*} - N^*)^2$.
The first two terms on the right hand side of Equation~\ref{eq:start-kappa}, thus obtained, are shown in Figure~\ref{fig:harmonic}d.
In the sparse sampling approach, the final term, $F_{\kappa, N^*}$, is obtained using thermodynamic integration.
The estimation of $F_{\kappa, N^*}$ is simplified somewhat if a single $\kappa$-value is adopted, and $N^*$ is varied in the biased simulations.
In particular, by using $dF_{\kappa, N^*} / dN^* = \langle dU_{\kappa, N^*} / dN^* \rangle_{\kappa,N^*}$, 
and performing thermodynamic integration from a reference value, $N^*_{\rm ref}$, to the $N^*$ of interest,
$F_{N^*} \equiv F_{\kappa, N^*} - F_{\kappa, N^*_{\rm ref}}$ can be obtained as: 
\begin{equation}
F_{N^*} = 
\int_{N^*_{\rm ref}}^{N^*} \langle dU_{\kappa, {N^*}^{\prime}} / d{N^*}^{\prime} \rangle_{\kappa, {N^*}^{\prime}} \mathrm{d}{N^*}^{\prime} =
 \int_{N^*_{\rm ref}}^{N^*} \kappa ( {N^*}^{\prime} - \avgNt_{\kappa, {N^*}^{\prime}} ) \mathrm{d}{N^*}^{\prime}
\label{eq:fnstar}
\end{equation}
In this way, $F_{\kappa, N^*} = F_{N^*} + F_{\kappa, N^*_{\rm ref}}$ is obtained to within an unknown constant, $F_{\kappa, N^*_{\rm ref}}$, 
by using estimates of $\avgNt_{\kappa, N^*}$ obtained from the biased simulations.
The constant, $F_{\kappa, N^*_{\rm ref}}$, if desired, can be obtained by thermodynamic integration from 0 to $\kappa$ as: 
\begin{equation}
F_{\kappa, N^*_{\rm ref}} = 
\int_0^\kappa \langle 	dU_{\kappa^{\prime}, N^*_{\rm ref}} / d{\kappa}^{\prime}	 \rangle_{\kappa^{\prime}, N^*_{\rm ref}} \mathrm{d}\kappa^{\prime} =
\frac{1}{2} \int_0^\kappa \langle (\Nt - N^*_{\rm ref})^2 \rangle_{\kappa^{\prime}, N^*_{\rm ref}} \mathrm{d}\kappa^{\prime}. 
\label{eq:fkappa}
\end{equation}
Alternatively, if $N^*_{\rm ref}$ is chosen to be a basin of $\fvnt$, there will be significant overlap between the unbiased and the (reference) biased ensembles, allowing $F_{\kappa, N^*_{\rm ref}}$ to be estimated using free energy perturbation~\cite{good_practices}.

For system 2, we first plot the thermodynamic force (the integrand in Equation~\ref{eq:fnstar}) as a function of $N^*$ in Figure~\ref{fig:harmonic}e.
We note that the integrand varies non-monotonically with $N^*$ in contrast with the corresponding integrand for a linear biasing potential (Figure~\ref{fig:caveat}b).
We then choose $N^*_{\rm ref}$ to be $n^{(1)}_{\rm liq}$ (which is a minimum of $F^{(2)}_v(\tilde N))$, 
and use numerical integration to obtain $F_{N^*}$ for all the simulated $N^*$-values (Figure~\ref{fig:harmonic}f).
Having obtained $F_{N^*}$, we can then use Equation~\ref{eq:start-kappa} to obtain sparse sampling estimates of $\fvnt$ at $\tilde N = \avgNt_{\kappa, N^*}$ to within the constant, $F_{\kappa, N^*_{\rm ref}}$. 
We do not estimate $F_{\kappa, N^*_{\rm ref}}$, but instead shift $\fvnt$ vertically to set the zero of $\fvnt$ at $\tilde N = N^*_{\rm ref}$.
As shown in Figure~\ref{fig:harmonic}g, the free energy profile, $\fvnt$, thus obtained, 
is in excellent agreement with $F_v^{\rm (2)}(\tilde N)$, highlighting that when the sparse sampling method 
is used in conjunction with a parabolic potential, it can be used to characterize the free energetics of systems near coexistence.
A discussion of the similarities and differences between the sparse sampling method and related techniques, such as Umbrella Integration (UI)~\cite{Kastner:JCP:2005}, is included as supplementary information.

\subsection{Choosing the Biasing Potential Parameters Judiciously}
%
The sparse sampling method can be used to efficiently sample $\fvnt$;
however, its practical implementation relies on the judicious choice of the biasing potential parameters, $\kappa$ and $N^*$. 
As shown in Figure~\ref{fig:harmonic}e, the average thermodynamic force, $dF_{N^*}/dN^*$, does not vary monotonically with $N^*$;
thus, care must be exercised in choosing $N^*$-values that capture its functional form.
To do so, we propose choosing $N^*$-values in an adaptive fashion.
To begin with, we employ equally spaced $N^*$-values, which span the desired $\tilde N$-range (from 0 to $n_{\rm{liq}}$).
The average thermodynamic force, $dF_{N^*}/dN^* = \langle dU_{\kappa,N^*}/dN^* \rangle_{\kappa,N^*} = \kappa(N^* - \avgNt_{\kappa,N^*})$, 
obtained from the biased simulations, is then inspected,
and in regions where it changes relatively abruptly, 
$N^*$-values are chosen to perform additional simulations.
In section 3, we will illustrate this strategy to estimate $\fvnt$ in a large spherical volume in bulk water.

\begin{figure}
\begin{center}
\includegraphics[width=0.6\textwidth]{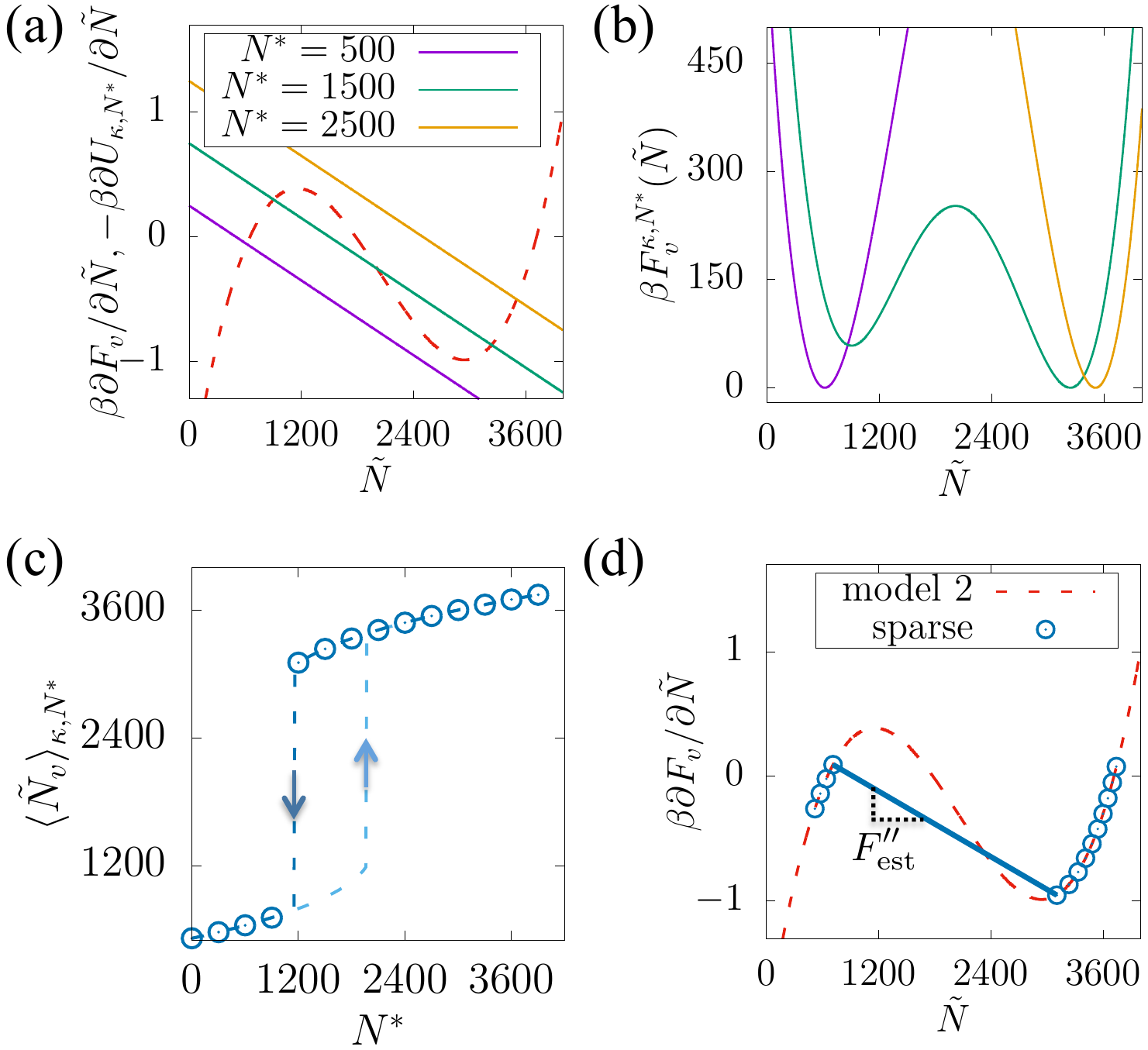}
\caption{
The importance of choosing a sufficiently large value of $\kappa$.
(a) When $\kappa$ is chosen to be too small, 
$-dU_{\kappa,N^*}/d\tilde N = -\kappa (\tilde N - N^*)$ intersects $\partial F_v^{(2)} / \partial \tilde N$ three times for certain $N^*$-values, 
suggesting that the corresponding biased free energy profiles, $F_v^{\kappa, N^*}(\tilde N)$, ought to be bistable.
(b) The biased free energy profiles, $F_v^{\kappa, N^*}(\tilde N)$, are indeed bistable for select $N^*$-values, as expected.
(c) In biased ensembles with bistable free energetics, $F_v^{\kappa, N^*}(\tilde N)$, regions with $d^2 F_v^{\kappa, N^*} / d\tilde N^2 < 0$ can not be sampled.
The sharp increase in $\avgNt_{\kappa, N^*}$ with increasing $N^*$, shown here, is symptomatic of this inability of biased simulations to sample certain $\Nt$-values. 
The bistability in $F_v^{\kappa, N^*}(\tilde N)$ also leads to hysteresis in $\avgNt_{\kappa, N^*}$; 
that is, $\avgNt_{\kappa, N^*}$ depends on whether the system was initialized in the dry state (upward arrow) or the wet state (downward arrow). 
(d) Estimates of $d F_v / d\tilde N$ can be obtained at $\tilde N = \avgNt_{\kappa, N^*}$ (blue circles) from the biased simulations;
the accuracy of these estimates is evident from their agreement with the exact $d F_v^{(2)} / d\tilde N$ (red dashed line).
A line connecting the points that correspond to the sharpest decrease in $d F_v / d\tilde N$ is shown (blue), and the magnitude of its slope is defined as $F''_{\rm est}$.
$F''_{\rm est}$ provides a lower bound (and a rough estimate) for $F''_{\rm max}$, the highest value that $-d^2 F_v / d\tilde N^2$ takes, 
and also informs our revised choice of $\kappa$.
}
\label{fig:choice}
\end{center}
\end{figure}

Care must also be exercised in our choice of $\kappa$.
In Figures~\ref{fig:harmonic}a and~\ref{fig:harmonic}b, we showed that when a parabolic potential with a sufficiently large $\kappa$ is used,  
 $d F_v/d\tilde N = -\kappa (\tilde N - N^*)$ has only one solution for all choices of $N^*$, 
 which then implies that $d F_v^{\kappa,N^*}/d\tilde N = 0$ has only one solution,
and the biased free energetics, $F_v^{\kappa,N^*}(\tilde N)$, are monostable for all $N^*$.
However, if $\kappa$ is chosen to be too small, $d F_v / d\tilde N = -\kappa (\tilde N - N^*)$ can have 3 roots for certain values of $N^*$, with the corresponding $F_v^{\kappa,N^*}(\tilde N)$ being bistable, as shown in Figures~\ref{fig:choice}a and~\ref{fig:choice}b (green) for $\beta\kappa = 0.0005$.
To ensure that $F_v^{\kappa,N^*}(\tilde N)$ is monostable for all $N^*$-values, $\kappa$, the slope of the lines in Figure~\ref{fig:choice}a, must be larger in magnitude than the largest negative slope of $d F_v/d\tilde N$.
In other words, a good choice of $\kappa$ is one that ensures the convexity of $F_v^{\kappa,N^*}(\tilde N)$ for all $\tilde N$ by satisfying the criterion:
\begin{equation}
\frac{ d^2 F_v^{\kappa,N^*} }{ d\tilde N^2 } = \frac{ d^2 F_v }{ d\tilde N^2} + \frac{ d^2 U_{\kappa, N^*} }{ d\tilde N^2} = \frac{ d^2 F_v }{ d\tilde N^2 } + \kappa > 0. 
\label{eq:kappa-crit}
\end{equation}
Thus, if $F''_{\rm max}$ is the largest value that $-d^2 F_v^{\kappa,N^*}/d\tilde N^2$ takes, we ought to choose $\kappa > F''_{\rm max}$.
However, given that we wish to estimate $F_v(\tilde N)$, the values taken by $d^2 F_v/d\tilde N^2$ are not known to us a priori, and neither is $F''_{\rm max}$.
Moreover, choosing excessively large values of $\kappa$ is also not recommended,
because they can lead to significant errors in estimates of the average thermodynamic force, $\kappa (N^* - \avgNt_{\kappa,N^*} )$, 
by amplifying small errors in our estimates of $\avgNt_{\kappa,N^*}$~\cite{zhu2012convergence}.
Such errors in the thermodynamic force are then propagated to $F_{N^*}$ through thermodynamic integration (Equation~\ref{eq:fnstar}), 
and eventually onto $\fvnt$ (Equation~\ref{eq:start-kappa}).
Thus, although it is important to choose a sufficiently large $\kappa$-value that exceeds $F''_{\rm max}$, 
it is not advisable to choose an excessively large $\kappa$-value 
because the length of time for which biased simulations must be run to obtain comparable errors in $\fvnt$ grows with $\kappa^2$.

Given these competing requirements, how do we optimally choose $\kappa$?
As a rule of thumb, we propose choosing $\beta\kappa$ to be a multiple of $\varNt_0^{-1}$, that is, $\beta\kappa = \alpha \varNt_0^{-1}$ with $\alpha$ in the range, $3 \le \alpha \le 5$.
The choice is underpinned by the observation that $\beta F''_{\rm max}$, the maximum negative value assumed by $\beta d^2 F_v/d\tilde N^2$, 
is often comparable to its (positive) value in the liquid basin, $\varNt_0^{-1}$, which is readily accessible from an unbiased simulation.
We have found that $\kappa$ chosen in accordance with this rule of thumb tends to satisfy the criterion in Equation~\ref{eq:kappa-crit} (that is, $\kappa > F''_{\rm max}$).
However, in the unlikely event that our initial choice of $\kappa$ is too small,
our biased simulations will exhibit a number of symptoms.
In particular, if $\kappa < F''_{\rm max}$, for certain $N^*$-values, $d^2 F_v^{\kappa,N^*}/d\tilde N^2$ will be negative
over a range of $\tilde N$-values, which will not be sampled in the biased simulations.
Consequently, there will be a sharp change in $\avgNt_{\kappa, N^*}$ as a function of $N^*$, 
as shown in Figure~\ref{fig:choice}c.
A second symptom of a small $\kappa$ is the appearance of hysteresis in $\avgNt_{\kappa, N^*}$ (Figure~\ref{fig:choice}c),
making it prudent to perform two sets biased simulations that are initialized in the liquid and vapor basins, respectively.
Thus, when $\kappa$ is chosen to be smaller than $F''_{\rm max}$, 
we encounter all of the same challenges that were encountered when using a linear biasing potential.

%
If our initial choice of $\kappa$ turns out to be too small, how do we choose a revised $\kappa$-value?
Can our initial set of biased simulations inform this revised choice?
To address these questions, we recognize that the response curves can also provide estimates of 
$d F_v / d\tilde N$ at $\tilde N = \avgNt_{\kappa,N^*}$ through:
\begin{equation}
\frac{d F_v }{  d\tilde N } (\tilde N = \avgNt_{\kappa,N^*}) \approx - \kappa (\avgNt_{\kappa,N^*} - N^*) 
\label{eq:kappa-deriv}
\end{equation}
To obtain this equation, we recognize that $d F_v / d\tilde N = d F_v^{\kappa,N^*} / d\tilde N - \kappa (\tilde N - N^*)$ (Equation~\ref{eq:start-kappa}), 
and that $d F_v^{\kappa,N^*} / d\tilde N$ ought to be 0 for $\tilde N = \avgNt_{\kappa,N^*}$.
Estimates of $d F_v / d\tilde N$ thus obtained are shown in Figure~\ref{fig:choice}d, and display two clusters at low and high $\tilde N$-values.
Although $d F_v / d\tilde N$ increases with increasing $\tilde N$ within the clusters, it decreases as we move from the low-$\tilde N$ cluster to the high-$\tilde N$ cluster, implying that $d^2 F_v / d\tilde N^2 < 0$ for intermediate $\tilde N$-values.
By identifying points that display the sharpest decrease in $d F_v / d\tilde N$,
and estimating the magnitude of the slope of the line connecting those points, $F''_{\rm est}$ (Figure~\ref{fig:choice}d), 
we can obtain a rough estimate of $F''_{\rm max}$; 
in fact, $F''_{\rm est}$ provides a lower bound on $F''_{\rm max}$, that is, $F''_{\rm est} \lesssim F''_{\rm max}$.
Thus, if our choice of $\kappa$ is smaller than $F''_{\rm est}$ --- that is, if $\kappa < F''_{\rm est} \lesssim F''_{\rm max}$ --- it was clearly too small.
To be safe, we propose the heuristic that $\kappa$ be larger than some multiple of $F''_{\rm est}$; 
that is, $\kappa > \alpha F''_{\rm est}$, with $3 \le \alpha \le 5$.
If this criterion is violated and either of the two low-$\kappa$ symptoms, discussed above, are observed,
we recommend that the revised $\kappa$ be chosen according to: $\kappa_{\rm new} = \alpha~\rm{max}(\kappa_{\rm old},F''_{\rm est})$;
that is, the revised (larger) value of $\kappa$ ought to be chosen to be $\alpha$ times either the previous (smaller) value of $\kappa$ or $F''_{\rm est}$, whichever is greater.
Thus, inspecting $d F_v / d\tilde N$ not only provides us with a third symptom of a small $\kappa$, 
but it also provides a way to choose the revised (higher) value of $\kappa$. 
Simulations with this higher value of $\kappa$ can then be repeated at all the $N^*$-values.
However, using a single $\kappa$-value for all biased simulations is not necessary.
We can also combine our initial low-$\kappa$ simulations (that are outside the hysteresis region) with
the newer high-$\kappa$ simulations to obtain $\fvnt$.
In section 4, we will illustrate how to integrate simulations with different $\kappa$-values. 

\section{Fluctuations in a Large Volume in Bulk Water}
\begin{figure}[htbp]
\begin{center}
\includegraphics[width=0.9\textwidth]{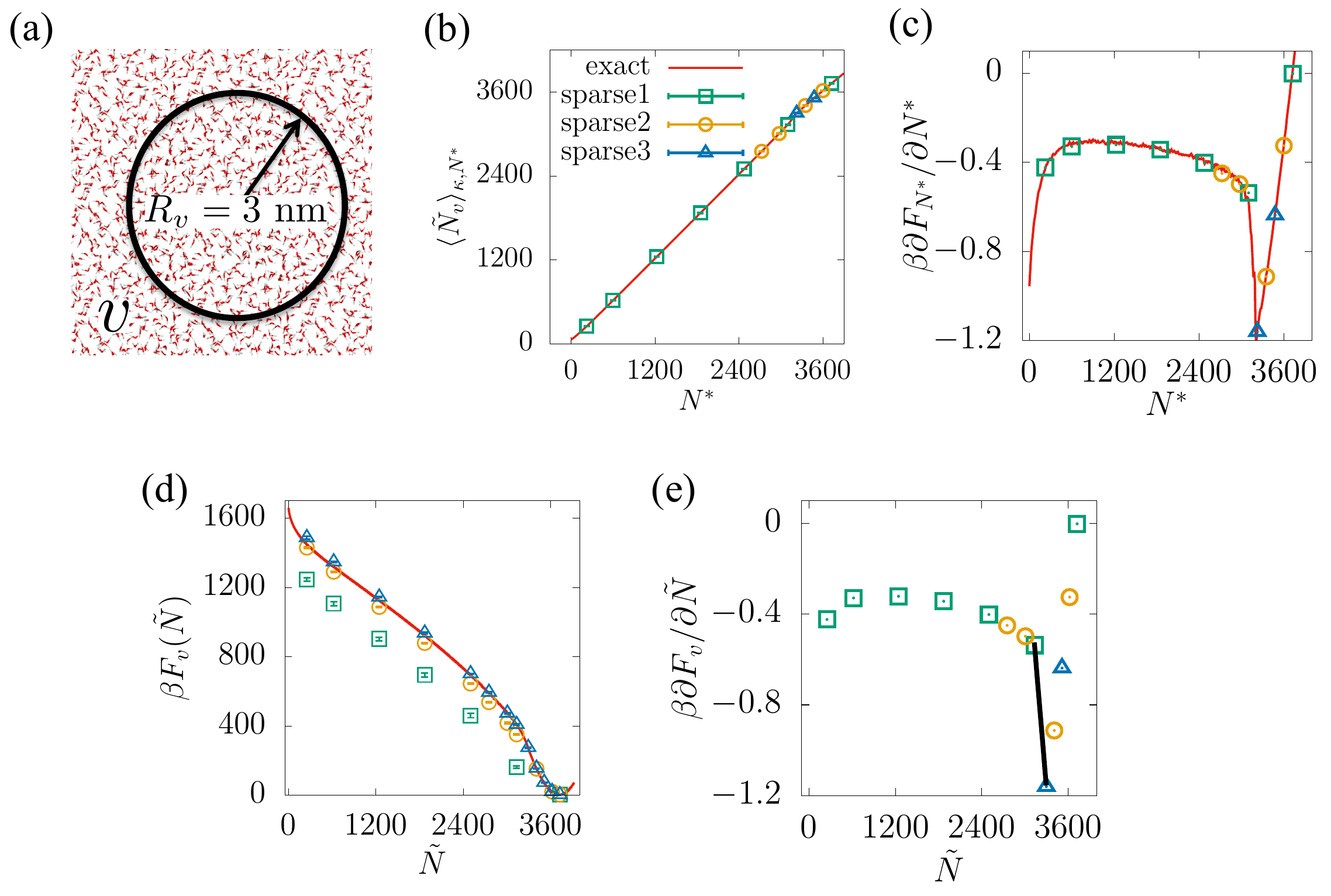}
\vspace{-0.15in}
\caption{
Employing the sparse sampling method to characterize the free energetics of water density fluctuations in a large volume in bulk water.
(a) A snapshot of the simulation box highlighting the spherical observation volume of radius, 
$R_v = 3$~ nm, in bulk water, which contains roughly 3700 water molecules on average.
(b) The variation of $\avgNt_{\kappa, N^*}$ with $N^*$ is roughly linear, facilitating the sampling of the entire $\Nt$-range of interest.
Green squares correspond to the data obtained from the 7 initial simulations, 
whereas the yellow circles correspond to results from 4 additional biased simulations, 
which were subsequently performed to better estimate the functional form of $dF_{N^*}/dN^*$.
Similarly, the blue triangles represent results from 2 more biased simulations that were performed in the third and final iteration. 
(c) Estimates of $dF_{N^*}/dN^*$ obtained from our biased simulations highlight its complex dependence on $N^*$.
Our initial simulations (green) suggest that at low $N^*$, $dF_{N^*}/dN^*$ increases as with increasing $N^*$, then decreases gradually, 
before increasing sharply at high $N^*$-values.
To better capture the functional form of $dF_{N^*}/dN^*$ at high $N^*$, we add two simulations each between the 3 highest $N^*$ values (yellow),
which improve our knowledge of its functional form, and thereby ought to improve our estimates of $F_{N^*}$, and eventually, $\fvnt$.
%
This procedure of successively adding simulations in regions 
where $dF_{N^*}/dN^*$ varies sharply with $N^*$
can be continued; here we add two more simulations in the third iteration (blue).
(d) The sparse sampled $\fvnt$ obtained using only the 7 initial simulations (green) is in good qualitative agreement with the exact umbrella sampling result (red line). The targeted addition of 4 more simulations (yellow) results in quantitative agreement with umbrella sampling, but is achieved with only a few percent of the computational expense. Adding the final 2 simulations (blue) results in only a small change in our estimate of $\fvnt$.
(e) Estimates of $\partial F_v / \partial \tilde N$ obtained at $\tilde N = \avgNt_{\kappa, N^*}$ highlight a sharp decrease around $\tilde N = 3200$;
the magnitude, $F''_{\rm est}$, of the slope of the line (black) connecting the points around this decrease provides us with a rough estimate for $F''_{\rm max}$.
Given that our choice of $\kappa$ is sufficiently (4.2 times) larger than $F''_{\rm est}$, we do not anticipate bistability in our biased simulations.
}
\label{fig:sph3}
\end{center}
\end{figure}

%
Here we use sparse sampling to characterize the free energetics, $\fvnt$, of observing $\tilde N$ waters in a spherical volume of radius, $R_v = 3$~nm, in bulk water.
Because the volume contains a large number, $n_{\rm liq} \approx 3700$, water molecules on average (Figure~\ref{fig:sph3}a), 
a large number of simulations are needed to obtain $\fvnt$ using umbrella sampling.
Indeed, 147 biased simulations, each run for 1~ns, were needed to ensure overlap between adjacent biased ensembles, 
and obtain the free energy profile, $\fvnt$, shown in Figure~\ref{fig:sph3}d (red curve).
Sparse sampling provides an efficient alternative to obtain $\fvnt$ at sparsely sampled $\tilde N$-values;
given the proximity of bulk water at ambient conditions to liquid-vapor coexistence, we make use of a parabolic potential, $U_{\kappa,N^*}$.
We first perform an unbiased simulation to estimate, $\varNt_0$, and choose $\beta\kappa = \alpha \varNt_0^{-1} = 0.016$ using $\alpha =5$.
We then perform biased simulations with this value of $\kappa$, and 7 sparsely spaced $N^*$-values that span from 0 to $n^{(1)}_{\rm liq}$.
%

The average number of water molecules, $\avgNt_{\kappa, N^*}$, obtained from the 7 biased simulations increase as $N^*$ is increased (Figure~\ref{fig:sph3}b, green squares).
However, the thermodynamic force, $dF_{N^*}/dN^* = \kappa ( {N^*} - \avgNt_{\kappa, {N^*}} )$, which must be integrated to obtain $F_{N^*}$ (Equation~\ref{eq:fnstar}), varies non-monotonically with $N^*$, first increasing at low $N^*$, then decreasing gradually, only to increase abruptly at the highest $N^*$-values (Figure~\ref{fig:sph3}c).
This abrupt increase suggests that the functional form of the integrand may not be adequately captured in the high $N^*$-region.
Nevertheless, the sparse sampled $\fvnt$-values (Figure~\ref{fig:sph3}d, green squares), 
obtained using the $\avgNt_{\kappa, N^*}$-values from these 7 simulations alone, 
not only capture $\fvnt$ qualitatively, but also display semi-quantitative agreement 
with the exact $\fvnt$ obtained from umbrella sampling (red curve), 
with the error in the free energies being roughly 15\%. 

To better capture the functional form of $dF_{N^*}/dN^*$, and to improve the accuracy with which the $F_{N^*}$-values, and thereby the sparse sampled $\fvnt$ profile are estimated, we include four additional simulations; two each between the three highest $N^*$-values. 
This increases the resolution with which we are able to characterize $dF_{N^*}/dN^*$ in a targeted fashion, providing estimates in the $N^*$-range where they are needed the most.
The corresponding results, shown by the yellow circles in Figures~\ref{fig:sph3}b-d, highlight the improvement in our knowledge of the functional form of the $dF_{N^*}/dN^*$ (Figure~\ref{fig:sph3}c), and the corresponding improvement in our estimation of $\fvnt$ (Figure~\ref{fig:sph3}d), which result from the four additional simulations.
This procedure of inspecting the variation of the integrand with $N^*$, 
and employing additional simulations in regions of substantive variation
to adaptively augment our characterization of the functional form of the integrand  
can be repeated further to obtain even more accurate estimates of $\fvnt$.  
Here, we perform two more simulations at such $N^*$-values, 
and obtain estimates of $\fvnt$ (blue triangles), 
which are not only in excellent quantitative agreement with the umbrella sampling results,
but also differ only marginally from estimates in the previous iteration (yellow circles),
 suggesting that convergence has been achieved. 
However, in contrast with the umbrella sampling, which incurred a computational overhead of 147 ns, obtaining the sparse sampled $\fvnt$ (blue triangles) in Figure~\ref{fig:sph3}d required only 13 biased simulations, each run for 0.2~ns, for a total simulation time of 2.6~ns. 
Thus, we were able to estimate $\fvnt$ roughly 2 orders of magnitude faster using sparse sampling. 

Although no sharp jumps in $\avgNt_{\kappa,N^*}$ were observed upon increasing $N^*$ in Figure~\ref{fig:sph3}b, 
we nevertheless estimate $\partial F_v / \partial \tilde N$ at those $\tilde N = \avgNt_{\kappa,N^*}$-values (Figure~\ref{fig:sph3}e, symbols) 
to ensure that our choice of $\kappa$ was sufficiently large.
By connecting the points that display the sharpest decrease in $\partial F_v / \partial \tilde N$ with a line (Figure~\ref{fig:sph3}e, black line), 
we estimate $\beta F''_{\rm est} \approx 0.004$, and correspondingly, $\kappa / F''_{\rm est} \approx 4.2$, 
suggesting that our initial choice of $\kappa$ was sufficiently large.

\begin{figure}[htbp]
\begin{center}
\includegraphics[width=0.99\textwidth]{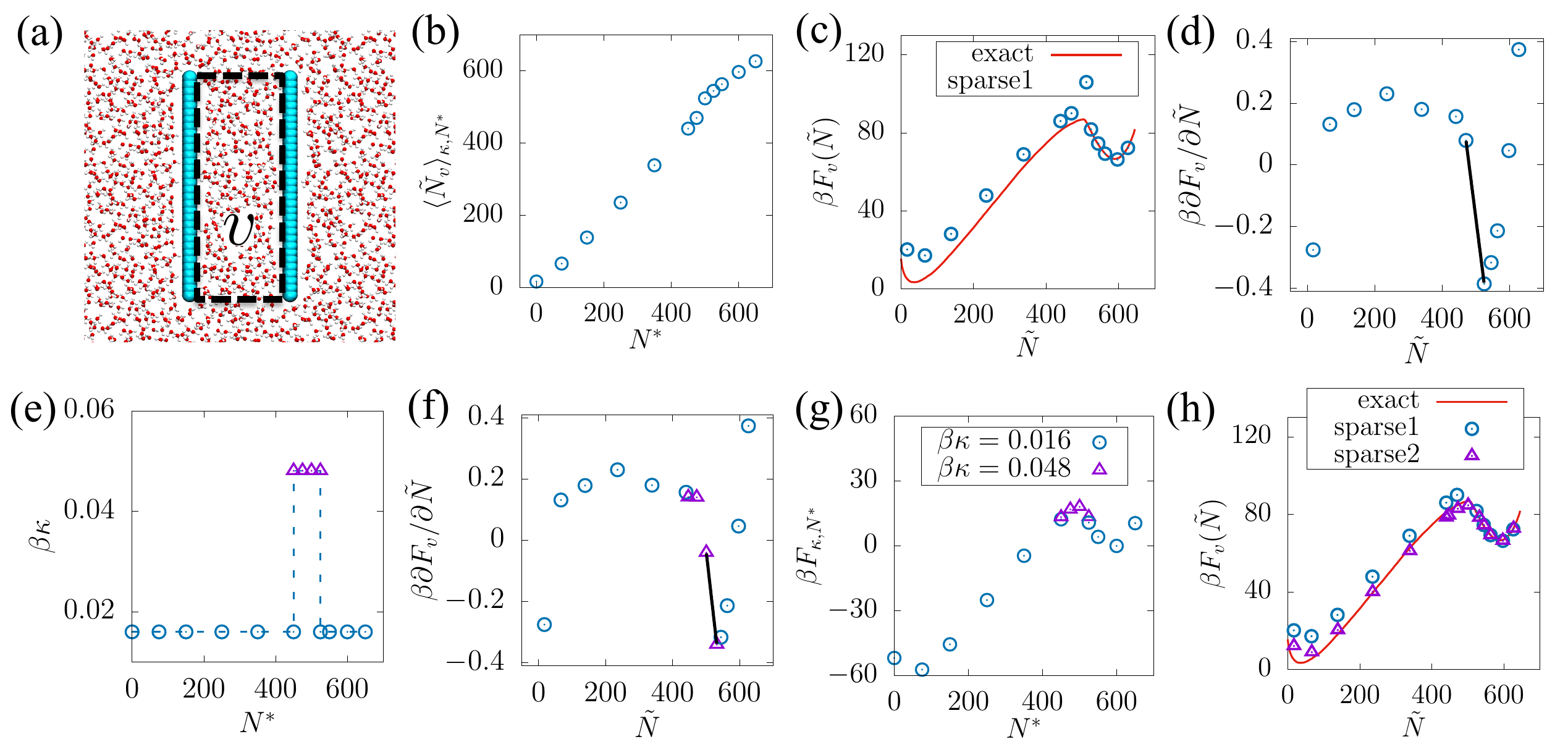}
\caption{
Employing sparse sampling to characterize $\fvnt$ in hydrophobic confinement.
(a) Snapshot of the simulation box highlighting water (red/white) in confinement between hydrophobic plates (cyan). 
The two hydrophobic plates, roughly 4~nm by 4~nm in size, are placed parallel to one another with a distance of 1.6~nm between them. 
The cuboid observation volume, $v$, which contains the confined waters, is also shown. 
(b) Estimates of $\avgNt_{\kappa,N^*}$ are shown for biased simulations performed using $\beta\kappa = 0.016$, and 12 sparsely spaced $N^*$-values.
(c) The exact $\fvnt$ obtained using umbrella sampling (red curve) displays two distinct basins, and a sharp change in slope (kink) near $\tilde N = 500$. 
The sparse sampling estimates of $\fvnt$ at $\tilde N = \avgNt_{\kappa,N^*}$ are in qualitative agreement with the umbrella sampling results;
however, there is a clear lack of quantitative agreement for $\tilde N < 500$.
(d) Estimates of $\partial F_v / \partial \tilde N$ obtained at $\tilde N = \avgNt_{\kappa, N^*}$ highlight a sharp decrease around $\tilde N = 500$;
the magnitude, $F''_{\rm est}$, of the corresponding slope (black line) provides us with a rough estimate of $F''_{\rm max}$.
(e) Based on the relative magnitudes of $\kappa$ and $F''_{\rm est}$, we elect to perform additional higher $\kappa$ simulations 
at select $N^*$-values in the vicinity of $\tilde N = 500$ (purple).
(f) Estimates of $\partial F_v / \partial \tilde N$ from the high-$\kappa$ simulations allow us to obtain a revised $F''_{\rm est}$; 
the revised $\kappa / F''_{\rm est} \approx 5$, suggesting that higher $\kappa$ ensembles are likely to be monostable. 
(g) Estimates of $F_{\kappa,N^*}$ for all the biased simulations are shown.
(h) The revised sparse sampling results are in excellent agreement with those obtained using umbrella sampling (exact). 
}
\label{fig:plates}
\end{center}
\end{figure}

\section{Water in hydrophobic confinement}
%
Here we use sparse sampling to characterize the free energetics of water density fluctuations in confinement 
between two square hydrophobic plates, roughly 4~nm by 4~nm in size, that are separated by 1.6~nm (Figure~\ref{fig:plates}a). 
For this separation, $\fvnt$ is known to feature two basins~\cite{Remsing:PNAS:2015}: a liquid (high $\tilde N$) basin and a vapor (low $\tilde N$) basin (Figure~\ref{fig:plates}c, red curve).
Moreover, in the vicinity of the barrier that separates the two basins, that is, near the maximum in $\fvnt$, there appears to be a ``kink'' or a sharp change in the slope of $\fvnt$.
In other words, over a small range of $\tilde N$-values, $\fvnt$ displays high negative curvature.
Recent work has attributed such a kink in $\fvnt$ to the presence of different dewetted morphologies on either side of the kink~\cite{Remsing:PNAS:2015,Prakash:PNAS:2016}.
In particular, Remsing {\it et al.} showed that
the low-$\tilde N$ (vapor) side of the kink features 
vapor tubes spanning the confined region between the two plates, 
whereas the high-$\tilde N$ (liquid) side of the kink features
isolated cavities that appear adjacent to one plate or the other~\cite{Remsing:PNAS:2015}. 
Similar behavior has also been reported for dewetting on nanotextured hydrophobic surfaces, wherein multiple kinks were observed in $\fvnt$, 
and were associated with transitions between different dewetted morphologies.~\cite{Prakash:PNAS:2016}

To sparse sample the system shown in Figure~\ref{fig:plates}a, we first choose 
$\beta\kappa = 0.016$, following our rule of thumb (section~2.5).
Given that this choice makes use of the curvature of $\fvnt$ in the liquid basin, 
it is unlikely to yield a $\kappa$-value that exceeds 
the large negative curvature in $\fvnt$ in the vicinity of a kink.
We thus recognize that our initial choice of $\kappa$ is likely to result in bistable biased free energy profiles, $F_v^{\kappa, N^*}(\tilde N)$;
however, because the large negative curvature in $\fvnt$ is localized to a small $\tilde N$-region near the kink, 
$F_v^{\kappa, N^*}(\tilde N)$ ought to be bistable for only a small range of $N^*$-values.
Using our initial choice of $\kappa$, we run biased simulations for 12 sparsely distributed $N^*$-values; 
the resulting values of $\avgNt_{\kappa,N^*}$ are shown in Figure~\ref{fig:plates}b as a function of $N^*$.
The increase in $\avgNt_{\kappa,N^*}$ with $N^*$ appears to be gradual, apart from a relatively sharp increase near $N^* = 500$.
By integrating $dF_{N^*}/dN^* = \kappa (N^* - \avgNt_{\kappa,N^*})$ with respect to $N^*$ to obtain estimates of $F_{N^*}$, 
and using Equation~\ref{eq:start-kappa}, we then obtain estimates of $\fvnt$ at $\tilde N = \avgNt_{\kappa,N^*}$ (Figure~\ref{fig:plates}c, blue circles).
The results display qualitative, but not quantitative agreement with the exact umbrella sampling results (Figure~\ref{fig:plates}c, red line).

To obtain more accurate estimates of $\fvnt$, and in particular, 
to ascertain whether the sharp increase in $\avgNt_{\kappa,N^*}$ near $N^* = 500$ 
is symptomatic of bistability in $F_v^{\kappa, N^*}(\tilde N)$,
we plot estimates of $d F_v / d\tilde N$ at $\tilde N = \avgNt_{\kappa,N^*}$ in Figure~\ref{fig:plates}d. 
We find that $d F_v / d\tilde N$ decreases sharply in the vicinity of $\tilde N = 500$, 
and that the magnitude of the corresponding slope, 
$\beta F''_{\rm est} \approx 0.0086 $,
and correspondingly, $\kappa / F''_{\rm est} \approx 1.86$, 
which is less than our chosen threshold of $\alpha = 3$. 
Taken together with the sharp increase in $\avgNt_{\kappa,N^*}$, 
this suggests that our initial choice of $\kappa$ was not large enough 
to prevent bistability in $F_v^{\kappa, N^*}(\tilde N)$ near $N^* = 500$.
Thus, either the biased simulations must be extended until
a converged $F_v^{\kappa, N^*}(\tilde N)$ is obtained with sufficient sampling of both its basins,
or a larger value of $\kappa$ ought to be chosen to ensure that the biased free energy profiles are monostable.
%
This is a non-trivial choice, which must take into consideration the possibility of hidden barriers, 
that is, barriers in collective variables that are orthogonal to $\Nt$, 
and the anticipated height of those barriers. 
Here we follow the prescription outlined in section~2.5, and choose a revised (higher) value of $\beta \kappa = 0.048$.
However, given that the sharp decrease in $d F_v / d\tilde N$ occurs over only a small range of $\tilde N$-values (Figure~\ref{fig:plates}d),
we do not repeat all the biased simulations using this higher value of $\kappa$.
Instead, we repeat our biased simulations with the higher $\kappa$-value only for $N^*$-values around 500;
the $\kappa$ and $N^*$-values that we adopt for the old and new biased simulations are shown in Figure~\ref{fig:plates}e. 
The higher $\kappa$ simulations allow us to augment our knowledge of $d F_v / d\tilde N$, 
and obtain a revised estimate for $F''_{\rm est}$ (Figure~\ref{fig:plates}f).
Comparing the revised values, 
$\beta F''_{\rm est} \approx 0.0096$,
and 
$\beta \kappa = 0.048$,
we get a revised estimate of $\kappa / F''_{\rm est} \approx 5$, 
which is above our threshold of $\alpha = 3$. 
Thus, we expect our revised choice of $\kappa$ to be large enough to facilitate sampling of $\Nt$ in the vicinity of the kink,
without the appearance of bistability in $F_v^{\kappa, N^*}(\tilde N)$ or hysteresis in $\avgNt_{\kappa,N^*}$.

Given that we no longer employ a single value of $\kappa$, 
in addition to estimating how $F_{\kappa, N^*}$ varies with $N^*$ (for a given $\kappa$),
we must also estimate the change in $F_{\kappa, N^*}$ as $\kappa$ is changed 
(for the two $N^*$ values shown in Figure~\ref{fig:plates}e).
We find that the latter can be readily estimated using the Bennett Acceptance Ratio (BAR) method~\cite{BAR}
because changing $\kappa$ (while keeping $N^*$ the same) 
leads to good overlap in $\Nt$-values sampled by the biased ensembles with different $\kappa$-values. 
In Figure~\ref{fig:plates}g, we plot estimates of $F_{\kappa,N^*}$ for all the biased ensembles considered. 
With all the $F_{\kappa,N^*}$-values in hand, Equation~\ref{eq:start-kappa} can then be used to obtain $\fvnt$ 
at $\tilde N = \avgNt_{\kappa,N^*}$; the resulting sparse sampled $\fvnt$-estimates are shown in Figure~\ref{fig:plates}h, 
and display good agreement with the umbrella sampling result.
Finally, we note that the total computational cost for performing sparse sampling is roughly 3 ns (16 simulations, 200~ps each), 
whereas the cost for performing umbrella sampling is roughly 230 ns (38 simulations, 6 ns each),
once again resulting in a nearly two orders of magnitude improvement in computational efficiency. 

\section{Conclusions}
%
A characterization of the free energetics of water density fluctuations in bulk water, at interfaces, and in hydrophobic confinement has 
substantially advanced our understanding of hydrophobic hydration, interactions, and assembly~\cite{Hummer:PNAS:1996,garde96,hummer98pnas,ashbaugh_SPT,Godawat:PNAS:2009,garde09prl,Patel:JPCB:2010,Rotenberg:JACS:2011,Patel:PNAS:2011,Garde:PNAS:2011,Remsing:JCP:2015,sanyal2016coarse}.
Such a characterization typically requires enhanced sampling techniques, such as umbrella sampling, which can be expensive~\cite{Patel:JPCB:2010,Patel:JSP:2011}.
In particular, the requirement that order parameter distributions obtained from adjacent biased simulations display overlap,
exacts a steep computational cost, and in certain contexts, can render umbrella sampling prohibitively expensive.
We recently introduced a sparse sampling method, which circumvents the overlap requirement central to umbrella sampling, and instead
uses thermodynamic integration to estimate free energy differences between the biased and unbiased ensembles~\cite{Xi:JCTC:2016}.
By employing only a few sparsely separated biased simulations, 
the sparse sampling method was able to estimate free energy profiles 
at a small fraction of the computational cost exacted by umbrella sampling.

Although the sparse sampling method, 
as introduced in ref.~\cite{Xi:JCTC:2016},
is efficient in its estimation of $\fvnt$,
it is only suitable for systems with convex free energy landscapes,
thereby excluding the important class of systems which are at or close to coexistence.
To estimate $\fvnt$ for systems near liquid-vapor coexistence, 
we generalized the sparse sampling method to sample free energy landscapes with concave regions.
Using model systems with analytical landscapes,
we first highlighted the challenges associated with sparse sampling systems near coexistence.
We then illustrated the use of a harmonic potential, $U_{\kappa,N^*}=\frac{1}{2}\kappa(\Nt - N^*)^2$, to regularize the free energy landscape.
Importantly, both the accuracy and efficiency of sparse sampling rely critically on the choice of the force constant, $\kappa$, 
as well as the $N^*$-values for which biased simulations are performed.
In order to minimize computational expense while ensuring reasonable accuracy,
we proposed the use of sensible heuristics which guide the initial choice of the biasing potential parameters, $\kappa$ and $N^*$.
We also introduced criteria for determining whether these initial choices needed to be refined, 
and proposed strategies for adaptively choosing new values of $\kappa$ and $N^*$, when needed.  
%

To illustrate the efficiency of the sparse sampling method, we studied two realistic systems close to liquid-vapor coexistence.
First, we characterized $\fvnt$ in a large volume containing roughly 3700 waters in bulk water at ambient conditions
Although liquid water is stable relative to its vapor at ambient conditions,
it can undergo a cavitation transition undergo sufficient tension,
which gives rise to concave features in $\fvnt$~\cite{zheng1991liquids,pallares2014anomalies,menzl2016molecular,biddle2017two}.
By choosing $N^*$-values adaptively, we obtained an accurate characterization of the free energetics of this process.
We also computed $\fvnt$ for water confined between two hydrophobic plates --- a system with distinct liquid and vapor basins.
Due to the presence of a small region of high negative curvature in $\fvnt$,
this system provides a more stringent test to the sparse sampling method.
By choosing $\kappa$-values adaptively, 
sparse sampling again facilitates an accurate characterization $\fvnt$.
In both cases, the computational expense of estimating $\fvnt$ is roughly two orders of magnitude smaller than umbrella sampling. 

Although we focus on the free energetics of water density fluctuations, 
the sparse sampling method itself is general, and can be readily used with any order parameter.
Moreover, it can also be extended to higher dimensions, and used to sample multiple order parameters.
The efficiency of the sparse sampling method makes it particularly attractive for characterizing the free energy landscapes 
of systems for which umbrella sampling is prohibitively expensive.
Such instances may arise when the (biased) simulations themselves are inherently expensive;
for example, it can be quite expensive to simulate crystalline systems with slow dynamics, 
or to explicitly account for polarizability or electronic effects~\cite{hassanali2013proton,remsing2014role,cisneros2016modeling,morawietz2016van}.
Alternatively, when a large number of biased simulations are needed to satisfy the overlap requirement, and
span the order parameter range of interest, umbrella sampling can again become very expensive; 
examples include the order parameter range of interest itself being large, inherent system fluctuations being small, 
and a high dimensional order parameter phase space~\cite{awasthi2016sampling,awasthi2017exploring,pfaendtner2015efficient}.
In each of these contexts, wherein performing extensive umbrella sampling simulations may not be feasible,
we hope that the estimation of free energy profiles will be facilitated by the sparse sampling method presented here.

\begin{acknowledgement}
The authors gratefully acknowledge financial support from the University of Pennsylvania Materials Research Science and Engineering Center (NSF UPENN MRSEC DMR 1720530), 
National Science Foundation grants CBET 1511437 and CBET 1652646, 
as well as the Charles E. Kaufman Foundation grant KA-2015-79204. 
%
A.J.P. is thankful to Talid Sinno for numerous helpful discussions.

\end{acknowledgement}

\noindent{\bf Simulation Details}
We perform all-atom molecular dynamics (MD) simulations using the GROMACS package~\cite{gmx4ref}, 
suitably modified to incorporate the biasing potentials of interest. 
The leap frog algorithm~\cite{Frenkel_Smit} was used to integrate the equations of motion with a 2 fs time-step, and
periodic boundary conditions were employed in all dimensions.
The SPC/E model of water~\cite{SPCE} was employed in all simulations, 
with the hydrogen bonds of water molecules being constrained using the SHAKE algorithm~\cite{SHAKE}.
The Particle Mesh Ewald algorithm~\cite{PME} was used to compute the long range electrostatic interactions,
and short range interactions Lennard-Jones and electrostatic interactions were truncated at 1~nm.
The canonical velocity-rescaling thermostat~\cite{V-Rescale} was used to maintain a system temperature of 300~K.
\\ {\bf Spherical Volume in Bulk Water} \\
The spherical observation volume with a radius, $R_v=3$~nm, was placed at the center of a cubic simulation box with a side of 10~nm. 
Simulations were performed in the NPT-ensemble at a pressure of 1~bar, which was maintained using the Parrinello-Rahman barostat~\cite{Parrinello-Rahman}.
To allow for equilibration, the first 100~ps were discarded from each biased simulation.
\\ {\bf Water in Hydrophobic Confinement} \\
The simulation setup for the hydrophobic plates is the same as that used in ref.~\cite{Remsing:PNAS:2015}. 
The plates are 4~nm by 4~nm in size and are separated by 1.6~nm, so that the confinement region is spanned by
 a $4\times4\times1.6$~nm$^3$ cuboid observation volume that is positioned between the plates. 
Simulations were performed in NVT-ensemble with a buffering water-vapor interface, which keeps the system at its coexistence pressure~\cite{Miller:PNAS:2007,Patel:JPCB:2010,Patel:JSP:2011}.
To allow for equilibration, the first 500~ps were discarded from each biased simulation.
%

\bibliography{Method}

\providecommand{\latin}[1]{#1}
\providecommand*\mcitethebibliography{\thebibliography}
\csname @ifundefined\endcsname{endmcitethebibliography}
  {\let\endmcitethebibliography\endthebibliography}{}
\begin{mcitethebibliography}{97}
\providecommand*\natexlab[1]{#1}
\providecommand*\mciteSetBstSublistMode[1]{}
\providecommand*\mciteSetBstMaxWidthForm[2]{}
\providecommand*\mciteBstWouldAddEndPuncttrue
  {\def\EndOfBibitem{\unskip.}}
\providecommand*\mciteBstWouldAddEndPunctfalse
  {\let\EndOfBibitem\relax}
\providecommand*\mciteSetBstMidEndSepPunct[3]{}
\providecommand*\mciteSetBstSublistLabelBeginEnd[3]{}
\providecommand*\EndOfBibitem{}
\mciteSetBstSublistMode{f}
\mciteSetBstMaxWidthForm{subitem}{(\alph{mcitesubitemcount})}
\mciteSetBstSublistLabelBeginEnd
  {\mcitemaxwidthsubitemform\space}
  {\relax}
  {\relax}

\bibitem[Stillinger(1973)]{stillinger}
Stillinger,~F.~H. \emph{J. Solution Chem.} \textbf{1973}, \emph{2},
  141--158\relax
\mciteBstWouldAddEndPuncttrue
\mciteSetBstMidEndSepPunct{\mcitedefaultmidpunct}
{\mcitedefaultendpunct}{\mcitedefaultseppunct}\relax
\EndOfBibitem
\bibitem[Lum \latin{et~al.}(1999)Lum, Chandler, and Weeks]{LCW}
Lum,~K.; Chandler,~D.; Weeks,~J.~D. \emph{J. Phys. Chem. B} \textbf{1999},
  \emph{103}, 4570--4577\relax
\mciteBstWouldAddEndPuncttrue
\mciteSetBstMidEndSepPunct{\mcitedefaultmidpunct}
{\mcitedefaultendpunct}{\mcitedefaultseppunct}\relax
\EndOfBibitem
\bibitem[Chandler(2005)]{Chandler:Nature:2005}
Chandler,~D. \emph{Nature} \textbf{2005}, \emph{437}, 640--647\relax
\mciteBstWouldAddEndPuncttrue
\mciteSetBstMidEndSepPunct{\mcitedefaultmidpunct}
{\mcitedefaultendpunct}{\mcitedefaultseppunct}\relax
\EndOfBibitem
\bibitem[Ashbaugh and Pratt(2006)Ashbaugh, and Pratt]{ashbaugh_SPT}
Ashbaugh,~H.~S.; Pratt,~L.~R. \emph{Rev. Mod. Phys.} \textbf{2006}, \emph{78},
  159\relax
\mciteBstWouldAddEndPuncttrue
\mciteSetBstMidEndSepPunct{\mcitedefaultmidpunct}
{\mcitedefaultendpunct}{\mcitedefaultseppunct}\relax
\EndOfBibitem
\bibitem[Varilly \latin{et~al.}(2011)Varilly, Patel, and Chandler]{LLCW}
Varilly,~P.; Patel,~A.~J.; Chandler,~D. \emph{J. Chem. Phys.} \textbf{2011},
  \emph{134}, 074109\relax
\mciteBstWouldAddEndPuncttrue
\mciteSetBstMidEndSepPunct{\mcitedefaultmidpunct}
{\mcitedefaultendpunct}{\mcitedefaultseppunct}\relax
\EndOfBibitem
\bibitem[Cerdeirina \latin{et~al.}(2011)Cerdeirina, Debenedetti, Rossky, and
  Giovambattista]{pgd:JPCL:2011}
Cerdeirina,~C.~A.; Debenedetti,~P.~G.; Rossky,~P.~J.; Giovambattista,~N.
  \emph{J. Phys. Chem. Lett.} \textbf{2011}, \emph{2}, 1000--1003\relax
\mciteBstWouldAddEndPuncttrue
\mciteSetBstMidEndSepPunct{\mcitedefaultmidpunct}
{\mcitedefaultendpunct}{\mcitedefaultseppunct}\relax
\EndOfBibitem
\bibitem[Vaikuntanathan and Geissler(2014)Vaikuntanathan, and
  Geissler]{Suri:PRL:2014}
Vaikuntanathan,~S.; Geissler,~P.~L. \emph{Phys. Rev. Lett.} \textbf{2014},
  \emph{112}, 020603\relax
\mciteBstWouldAddEndPuncttrue
\mciteSetBstMidEndSepPunct{\mcitedefaultmidpunct}
{\mcitedefaultendpunct}{\mcitedefaultseppunct}\relax
\EndOfBibitem
\bibitem[Vaikuntanathan \latin{et~al.}(2016)Vaikuntanathan, Rotskoff, Hudson,
  and Geissler]{Suri:PNAS:2016}
Vaikuntanathan,~S.; Rotskoff,~G.; Hudson,~A.; Geissler,~P.~L. \emph{Proc. Natl.
  Acad. Sci. U.S.A.} \textbf{2016}, \emph{113}, E2224--E2230\relax
\mciteBstWouldAddEndPuncttrue
\mciteSetBstMidEndSepPunct{\mcitedefaultmidpunct}
{\mcitedefaultendpunct}{\mcitedefaultseppunct}\relax
\EndOfBibitem
\bibitem[Xi and Patel(2016)Xi, and Patel]{Xi:PNAS:2016}
Xi,~E.; Patel,~A.~J. \emph{Proc. Natl. Acad. Sci. U.S.A.} \textbf{2016},
  \emph{113}, 4549--4551\relax
\mciteBstWouldAddEndPuncttrue
\mciteSetBstMidEndSepPunct{\mcitedefaultmidpunct}
{\mcitedefaultendpunct}{\mcitedefaultseppunct}\relax
\EndOfBibitem
\bibitem[Desgranges and Delhommelle(2016)Desgranges, and
  Delhommelle]{desgranges2016free}
Desgranges,~C.; Delhommelle,~J. \emph{The Journal of chemical physics}
  \textbf{2016}, \emph{145}, 204112\relax
\mciteBstWouldAddEndPuncttrue
\mciteSetBstMidEndSepPunct{\mcitedefaultmidpunct}
{\mcitedefaultendpunct}{\mcitedefaultseppunct}\relax
\EndOfBibitem
\bibitem[Lee \latin{et~al.}(1984)Lee, McCammon, and Rossky]{rossky84}
Lee,~C.-Y.; McCammon,~J.~A.; Rossky,~P.~J. \emph{J. Chem. Phys.} \textbf{1984},
  \emph{80}, 4448--4455\relax
\mciteBstWouldAddEndPuncttrue
\mciteSetBstMidEndSepPunct{\mcitedefaultmidpunct}
{\mcitedefaultendpunct}{\mcitedefaultseppunct}\relax
\EndOfBibitem
\bibitem[Giovambattista \latin{et~al.}(2007)Giovambattista, Debenedetti, and
  Rossky]{pgd:JPCB:2007}
Giovambattista,~N.; Debenedetti,~P.~G.; Rossky,~P.~J. \emph{J. Phys. Chem. B}
  \textbf{2007}, \emph{111}, 9581--9587\relax
\mciteBstWouldAddEndPuncttrue
\mciteSetBstMidEndSepPunct{\mcitedefaultmidpunct}
{\mcitedefaultendpunct}{\mcitedefaultseppunct}\relax
\EndOfBibitem
\bibitem[Mittal and Hummer(2008)Mittal, and Hummer]{mittal_pnas08}
Mittal,~J.; Hummer,~G. \emph{Proc. Natl. Acad. Sci. U.S.A.} \textbf{2008},
  \emph{105}, 20130--20135\relax
\mciteBstWouldAddEndPuncttrue
\mciteSetBstMidEndSepPunct{\mcitedefaultmidpunct}
{\mcitedefaultendpunct}{\mcitedefaultseppunct}\relax
\EndOfBibitem
\bibitem[Godawat \latin{et~al.}(2009)Godawat, Jamadagni, and
  Garde]{Godawat:PNAS:2009}
Godawat,~R.; Jamadagni,~S.~N.; Garde,~S. \emph{Proc. Natl. Acad. Sci. U.S.A.}
  \textbf{2009}, \emph{106}, 15119 -- 15124\relax
\mciteBstWouldAddEndPuncttrue
\mciteSetBstMidEndSepPunct{\mcitedefaultmidpunct}
{\mcitedefaultendpunct}{\mcitedefaultseppunct}\relax
\EndOfBibitem
\bibitem[Giovambattista \latin{et~al.}(2009)Giovambattista, Debenedetti, and
  Rossky]{Giovambattista:PNAS:2009}
Giovambattista,~N.; Debenedetti,~P.~G.; Rossky,~P.~J. \emph{Proc. Natl. Acad.
  Sci. U.S.A.} \textbf{2009}, \emph{106}, 15181--15185\relax
\mciteBstWouldAddEndPuncttrue
\mciteSetBstMidEndSepPunct{\mcitedefaultmidpunct}
{\mcitedefaultendpunct}{\mcitedefaultseppunct}\relax
\EndOfBibitem
\bibitem[Patel \latin{et~al.}(2010)Patel, Varilly, and
  Chandler]{Patel:JPCB:2010}
Patel,~A.~J.; Varilly,~P.; Chandler,~D. \emph{J. Phys. Chem. B} \textbf{2010},
  \emph{114}, 1632 -- 1637\relax
\mciteBstWouldAddEndPuncttrue
\mciteSetBstMidEndSepPunct{\mcitedefaultmidpunct}
{\mcitedefaultendpunct}{\mcitedefaultseppunct}\relax
\EndOfBibitem
\bibitem[Rotenberg \latin{et~al.}(2011)Rotenberg, Patel, and
  Chandler]{Rotenberg:JACS:2011}
Rotenberg,~B.; Patel,~A.~J.; Chandler,~D. \emph{J. Am. Chem. Soc.}
  \textbf{2011}, \emph{133}, 20521 -- 20527\relax
\mciteBstWouldAddEndPuncttrue
\mciteSetBstMidEndSepPunct{\mcitedefaultmidpunct}
{\mcitedefaultendpunct}{\mcitedefaultseppunct}\relax
\EndOfBibitem
\bibitem[Patel \latin{et~al.}(2011)Patel, Varilly, Jamadagni, Acharya, Garde,
  and Chandler]{Patel:PNAS:2011}
Patel,~A.~J.; Varilly,~P.; Jamadagni,~S.~N.; Acharya,~H.; Garde,~S.;
  Chandler,~D. \emph{Proc. Natl. Acad. Sci. U.S.A.} \textbf{2011}, \emph{108},
  17678 -- 17683\relax
\mciteBstWouldAddEndPuncttrue
\mciteSetBstMidEndSepPunct{\mcitedefaultmidpunct}
{\mcitedefaultendpunct}{\mcitedefaultseppunct}\relax
\EndOfBibitem
\bibitem[Patel \latin{et~al.}(2012)Patel, Varilly, Jamadagni, Hagan, Chandler,
  and Garde]{Patel:JPCB:2012}
Patel,~A.~J.; Varilly,~P.; Jamadagni,~S.~N.; Hagan,~M.~F.; Chandler,~D.;
  Garde,~S. \emph{J. Phys. Chem. B} \textbf{2012}, \emph{116}, 2498 --
  2503\relax
\mciteBstWouldAddEndPuncttrue
\mciteSetBstMidEndSepPunct{\mcitedefaultmidpunct}
{\mcitedefaultendpunct}{\mcitedefaultseppunct}\relax
\EndOfBibitem
\bibitem[Luzar and Leung(2000)Luzar, and Leung]{Luzar:JCP:2000}
Luzar,~A.; Leung,~K. \emph{J. Chem. Phys.} \textbf{2000}, \emph{113},
  5836--5844\relax
\mciteBstWouldAddEndPuncttrue
\mciteSetBstMidEndSepPunct{\mcitedefaultmidpunct}
{\mcitedefaultendpunct}{\mcitedefaultseppunct}\relax
\EndOfBibitem
\bibitem[Hummer \latin{et~al.}(2001)Hummer, Rasaiah, and
  Noworyta]{hummer_nanotube}
Hummer,~G.; Rasaiah,~J.~C.; Noworyta,~J.~P. \emph{Nature} \textbf{2001},
  \emph{414}, 188--190\relax
\mciteBstWouldAddEndPuncttrue
\mciteSetBstMidEndSepPunct{\mcitedefaultmidpunct}
{\mcitedefaultendpunct}{\mcitedefaultseppunct}\relax
\EndOfBibitem
\bibitem[Leung \latin{et~al.}(2003)Leung, Luzar, and Bratko]{Luzar:PRL:2003}
Leung,~K.; Luzar,~A.; Bratko,~D. \emph{Phys. Rev. Lett.} \textbf{2003},
  \emph{90}, 065502\relax
\mciteBstWouldAddEndPuncttrue
\mciteSetBstMidEndSepPunct{\mcitedefaultmidpunct}
{\mcitedefaultendpunct}{\mcitedefaultseppunct}\relax
\EndOfBibitem
\bibitem[Patankar(2004)]{Patankar:Langmuir:2004}
Patankar,~N.~A. \emph{Langmuir} \textbf{2004}, \emph{20}, 7097--7102\relax
\mciteBstWouldAddEndPuncttrue
\mciteSetBstMidEndSepPunct{\mcitedefaultmidpunct}
{\mcitedefaultendpunct}{\mcitedefaultseppunct}\relax
\EndOfBibitem
\bibitem[Bolhuis and Chandler(2000)Bolhuis, and Chandler]{bolhuis}
Bolhuis,~P.~G.; Chandler,~D. \emph{J. Chem. Phys.} \textbf{2000}, \emph{113},
  8154--8160\relax
\mciteBstWouldAddEndPuncttrue
\mciteSetBstMidEndSepPunct{\mcitedefaultmidpunct}
{\mcitedefaultendpunct}{\mcitedefaultseppunct}\relax
\EndOfBibitem
\bibitem[Giovambattista \latin{et~al.}(2006)Giovambattista, Rossky, and
  Debenedetti]{pgd06}
Giovambattista,~N.; Rossky,~P.~J.; Debenedetti,~P.~G. \emph{Phys. Rev. E}
  \textbf{2006}, \emph{73}, 041604\relax
\mciteBstWouldAddEndPuncttrue
\mciteSetBstMidEndSepPunct{\mcitedefaultmidpunct}
{\mcitedefaultendpunct}{\mcitedefaultseppunct}\relax
\EndOfBibitem
\bibitem[Giovambattista \latin{et~al.}(2007)Giovambattista, Debenedetti, and
  Rossky]{Giovambattista:JPCC:2007}
Giovambattista,~N.; Debenedetti,~P.~G.; Rossky,~P.~J. \emph{J. Phys. Chem. C}
  \textbf{2007}, \emph{111}, 1323--1332\relax
\mciteBstWouldAddEndPuncttrue
\mciteSetBstMidEndSepPunct{\mcitedefaultmidpunct}
{\mcitedefaultendpunct}{\mcitedefaultseppunct}\relax
\EndOfBibitem
\bibitem[Choudhury and Pettitt(2007)Choudhury, and Pettitt]{chou_dewet}
Choudhury,~N.; Pettitt,~B.~M. \emph{J. Am. Chem. Soc.} \textbf{2007},
  \emph{129}, 4847--4852\relax
\mciteBstWouldAddEndPuncttrue
\mciteSetBstMidEndSepPunct{\mcitedefaultmidpunct}
{\mcitedefaultendpunct}{\mcitedefaultseppunct}\relax
\EndOfBibitem
\bibitem[Rasaiah \latin{et~al.}(2008)Rasaiah, Garde, and
  Hummer]{rasaiah2008water}
Rasaiah,~J.~C.; Garde,~S.; Hummer,~G. \emph{Annu. Rev. Phys. Chem.}
  \textbf{2008}, \emph{59}, 713--740\relax
\mciteBstWouldAddEndPuncttrue
\mciteSetBstMidEndSepPunct{\mcitedefaultmidpunct}
{\mcitedefaultendpunct}{\mcitedefaultseppunct}\relax
\EndOfBibitem
\bibitem[Hua \latin{et~al.}(2009)Hua, Zangi, and Berne]{Berne:JPCC:2009}
Hua,~L.; Zangi,~R.; Berne,~B.~J. \emph{J. Phys. Chem. C} \textbf{2009},
  \emph{113}, 5244--5253\relax
\mciteBstWouldAddEndPuncttrue
\mciteSetBstMidEndSepPunct{\mcitedefaultmidpunct}
{\mcitedefaultendpunct}{\mcitedefaultseppunct}\relax
\EndOfBibitem
\bibitem[Berne \latin{et~al.}(2009)Berne, Weeks, and Zhou]{berne_rev09}
Berne,~B.~J.; Weeks,~J.~D.; Zhou,~R. \emph{Annu. Rev. Phys. Chem.}
  \textbf{2009}, \emph{60}, 85--103\relax
\mciteBstWouldAddEndPuncttrue
\mciteSetBstMidEndSepPunct{\mcitedefaultmidpunct}
{\mcitedefaultendpunct}{\mcitedefaultseppunct}\relax
\EndOfBibitem
\bibitem[Mittal and Hummer(2010)Mittal, and Hummer]{Mittal:Faraday:2010}
Mittal,~J.; Hummer,~G. \emph{Faraday Discuss.} \textbf{2010}, \emph{146},
  341--352\relax
\mciteBstWouldAddEndPuncttrue
\mciteSetBstMidEndSepPunct{\mcitedefaultmidpunct}
{\mcitedefaultendpunct}{\mcitedefaultseppunct}\relax
\EndOfBibitem
\bibitem[Wang \latin{et~al.}(2011)Wang, Bratko, and Luzar]{Wang:PNAS:2011}
Wang,~J.; Bratko,~D.; Luzar,~A. \emph{Proc. Natl. Acad. Sci. U.S.A.}
  \textbf{2011}, \emph{108}, 6374--6379\relax
\mciteBstWouldAddEndPuncttrue
\mciteSetBstMidEndSepPunct{\mcitedefaultmidpunct}
{\mcitedefaultendpunct}{\mcitedefaultseppunct}\relax
\EndOfBibitem
\bibitem[Giacomello \latin{et~al.}(2012)Giacomello, Chinappi, Meloni, and
  Casciola]{Giacomello:PRL:2012}
Giacomello,~A.; Chinappi,~M.; Meloni,~S.; Casciola,~C.~M. \emph{Phys. Rev.
  Lett.} \textbf{2012}, \emph{109}, 226102\relax
\mciteBstWouldAddEndPuncttrue
\mciteSetBstMidEndSepPunct{\mcitedefaultmidpunct}
{\mcitedefaultendpunct}{\mcitedefaultseppunct}\relax
\EndOfBibitem
\bibitem[Altabet and Debenedetti(2014)Altabet, and
  Debenedetti]{Altabet:JCP:2014}
Altabet,~Y.~E.; Debenedetti,~P.~G. \emph{J. Chem. Phys.} \textbf{2014},
  \emph{141}, 18C531\relax
\mciteBstWouldAddEndPuncttrue
\mciteSetBstMidEndSepPunct{\mcitedefaultmidpunct}
{\mcitedefaultendpunct}{\mcitedefaultseppunct}\relax
\EndOfBibitem
\bibitem[Remsing \latin{et~al.}(2015)Remsing, Xi, Vembanur, Sharma,
  Debenedetti, Garde, and Patel]{Remsing:PNAS:2015}
Remsing,~R.~C.; Xi,~E.; Vembanur,~S.; Sharma,~S.; Debenedetti,~P.~G.;
  Garde,~S.; Patel,~A.~J. \emph{Proc. Natl. Acad. Sci. U.S.A.} \textbf{2015},
  \emph{112}, 8181--8186\relax
\mciteBstWouldAddEndPuncttrue
\mciteSetBstMidEndSepPunct{\mcitedefaultmidpunct}
{\mcitedefaultendpunct}{\mcitedefaultseppunct}\relax
\EndOfBibitem
\bibitem[Prakash \latin{et~al.}(2016)Prakash, Xi, and Patel]{Prakash:PNAS:2016}
Prakash,~S.; Xi,~E.; Patel,~A.~J. \emph{Proc. Natl. Acad. Sci. U.S.A.}
  \textbf{2016}, \emph{113}, 5508 -- 5513\relax
\mciteBstWouldAddEndPuncttrue
\mciteSetBstMidEndSepPunct{\mcitedefaultmidpunct}
{\mcitedefaultendpunct}{\mcitedefaultseppunct}\relax
\EndOfBibitem
\bibitem[Altabet \latin{et~al.}(2017)Altabet, Haji-Akbari, and
  Debenedetti]{Altabet:PNAS:2017}
Altabet,~Y.~E.; Haji-Akbari,~A.; Debenedetti,~P.~G. \emph{Proc. Natl. Acad.
  Sci. U.S.A.} \textbf{2017}, \relax
\mciteBstWouldAddEndPunctfalse
\mciteSetBstMidEndSepPunct{\mcitedefaultmidpunct}
{}{\mcitedefaultseppunct}\relax
\EndOfBibitem
\bibitem[Chaimovich and Shell(2013)Chaimovich, and Shell]{chaimovich2013length}
Chaimovich,~A.; Shell,~M.~S. \emph{Physical Review E} \textbf{2013}, \emph{88},
  052313\relax
\mciteBstWouldAddEndPuncttrue
\mciteSetBstMidEndSepPunct{\mcitedefaultmidpunct}
{\mcitedefaultendpunct}{\mcitedefaultseppunct}\relax
\EndOfBibitem
\bibitem[Wang \latin{et~al.}(2017)Wang, Jenkins, McGinley, Sinno, and
  Crocker]{wang2017colloidal}
Wang,~Y.; Jenkins,~I.~C.; McGinley,~J.~T.; Sinno,~T.; Crocker,~J.~C.
  \emph{Nature Comm.} \textbf{2017}, \emph{8}, 14173\relax
\mciteBstWouldAddEndPuncttrue
\mciteSetBstMidEndSepPunct{\mcitedefaultmidpunct}
{\mcitedefaultendpunct}{\mcitedefaultseppunct}\relax
\EndOfBibitem
\bibitem[Zhou \latin{et~al.}(2004)Zhou, Huang, Margulis, and Berne]{berne04}
Zhou,~R.; Huang,~X.; Margulis,~C.~J.; Berne,~B.~J. \emph{Science}
  \textbf{2004}, \emph{305}, 1605--1609\relax
\mciteBstWouldAddEndPuncttrue
\mciteSetBstMidEndSepPunct{\mcitedefaultmidpunct}
{\mcitedefaultendpunct}{\mcitedefaultseppunct}\relax
\EndOfBibitem
\bibitem[Freed \latin{et~al.}(2011)Freed, Garde, and Cramer]{Freed:JPCB:2011}
Freed,~A.~S.; Garde,~S.; Cramer,~S.~M. \emph{J. Phys. Chem. B} \textbf{2011},
  \emph{115}, 13320--13327\relax
\mciteBstWouldAddEndPuncttrue
\mciteSetBstMidEndSepPunct{\mcitedefaultmidpunct}
{\mcitedefaultendpunct}{\mcitedefaultseppunct}\relax
\EndOfBibitem
\bibitem[Snyder \latin{et~al.}(2011)Snyder, Mecinovi{\'c}, Moustakas, Thomas,
  Harder, Mack, Lockett, H{\'e}roux, Sherman, and Whitesides]{Snyder:PNAS:2011}
Snyder,~P.~W.; Mecinovi{\'c},~J.; Moustakas,~D.~T.; Thomas,~S.~W.; Harder,~M.;
  Mack,~E.~T.; Lockett,~M.~R.; H{\'e}roux,~A.; Sherman,~W.; Whitesides,~G.~M.
  \emph{Proc. Natl. Acad. Sci. U.S.A.} \textbf{2011}, \emph{108},
  17889--17894\relax
\mciteBstWouldAddEndPuncttrue
\mciteSetBstMidEndSepPunct{\mcitedefaultmidpunct}
{\mcitedefaultendpunct}{\mcitedefaultseppunct}\relax
\EndOfBibitem
\bibitem[Wang \latin{et~al.}(2011)Wang, Berne, and
  Friesner]{Wang-Berne:PNAS:2011}
Wang,~L.; Berne,~B.~J.; Friesner,~R.~A. \emph{Proc. Natl. Acad. Sci. U.S.A.}
  \textbf{2011}, \emph{108}, 1326--1330\relax
\mciteBstWouldAddEndPuncttrue
\mciteSetBstMidEndSepPunct{\mcitedefaultmidpunct}
{\mcitedefaultendpunct}{\mcitedefaultseppunct}\relax
\EndOfBibitem
\bibitem[Shahraz \latin{et~al.}(2012)Shahraz, Borhan, and
  Fichthorn]{Shahraz:Langmuir:2012}
Shahraz,~A.; Borhan,~A.; Fichthorn,~K.~A. \emph{Langmuir} \textbf{2012},
  \emph{28}, 14227--14237\relax
\mciteBstWouldAddEndPuncttrue
\mciteSetBstMidEndSepPunct{\mcitedefaultmidpunct}
{\mcitedefaultendpunct}{\mcitedefaultseppunct}\relax
\EndOfBibitem
\bibitem[Checco \latin{et~al.}(2014)Checco, Ocko, Rahman, Black, Tasinkevych,
  Giacomello, and Dietrich]{Checco:PRL:2014}
Checco,~A.; Ocko,~B.~M.; Rahman,~A.; Black,~C.~T.; Tasinkevych,~M.;
  Giacomello,~A.; Dietrich,~S. \emph{Phys. Rev. Lett.} \textbf{2014},
  \emph{112}, 216101\relax
\mciteBstWouldAddEndPuncttrue
\mciteSetBstMidEndSepPunct{\mcitedefaultmidpunct}
{\mcitedefaultendpunct}{\mcitedefaultseppunct}\relax
\EndOfBibitem
\bibitem[Savoy and Escobedo(2012)Savoy, and Escobedo]{savoy2012molecular}
Savoy,~E.~S.; Escobedo,~F.~A. \emph{Langmuir} \textbf{2012}, \emph{28},
  3412--3419\relax
\mciteBstWouldAddEndPuncttrue
\mciteSetBstMidEndSepPunct{\mcitedefaultmidpunct}
{\mcitedefaultendpunct}{\mcitedefaultseppunct}\relax
\EndOfBibitem
\bibitem[Sharma and Debenedetti(2012)Sharma, and Debenedetti]{Sharma:PNAS:2012}
Sharma,~S.; Debenedetti,~P.~G. \emph{Proc. Natl. Acad. Sci. U.S.A.}
  \textbf{2012}, \emph{109}, 4365--4370\relax
\mciteBstWouldAddEndPuncttrue
\mciteSetBstMidEndSepPunct{\mcitedefaultmidpunct}
{\mcitedefaultendpunct}{\mcitedefaultseppunct}\relax
\EndOfBibitem
\bibitem[Sharma and Debenedetti(2012)Sharma, and Debenedetti]{Sharma:JPCB:2012}
Sharma,~S.; Debenedetti,~P.~G. \emph{J. Phys. Chem. B} \textbf{2012},
  \emph{116}, 13282--13289\relax
\mciteBstWouldAddEndPuncttrue
\mciteSetBstMidEndSepPunct{\mcitedefaultmidpunct}
{\mcitedefaultendpunct}{\mcitedefaultseppunct}\relax
\EndOfBibitem
\bibitem[Desgranges and Delhommelle(2017)Desgranges, and
  Delhommelle]{desgranges2017free}
Desgranges,~C.; Delhommelle,~J. \emph{The Journal of Chemical Physics}
  \textbf{2017}, \emph{146}, 184104\relax
\mciteBstWouldAddEndPuncttrue
\mciteSetBstMidEndSepPunct{\mcitedefaultmidpunct}
{\mcitedefaultendpunct}{\mcitedefaultseppunct}\relax
\EndOfBibitem
\bibitem[ten Wolde and Chandler(2002)ten Wolde, and
  Chandler]{ChandlerPolymerLattice}
ten Wolde,~P.~R.; Chandler,~D. \emph{Proc. Natl. Acad. Sci. U.S.A.}
  \textbf{2002}, \emph{99}, 6539--6543\relax
\mciteBstWouldAddEndPuncttrue
\mciteSetBstMidEndSepPunct{\mcitedefaultmidpunct}
{\mcitedefaultendpunct}{\mcitedefaultseppunct}\relax
\EndOfBibitem
\bibitem[Miller \latin{et~al.}(2007)Miller, Vanden-Eijnden, and
  Chandler]{Miller:PNAS:2007}
Miller,~T.; Vanden-Eijnden,~E.; Chandler,~D. \emph{Proc. Natl. Acad. Sci.
  U.S.A.} \textbf{2007}, \emph{104}, 14559--14564\relax
\mciteBstWouldAddEndPuncttrue
\mciteSetBstMidEndSepPunct{\mcitedefaultmidpunct}
{\mcitedefaultendpunct}{\mcitedefaultseppunct}\relax
\EndOfBibitem
\bibitem[Setny \latin{et~al.}(2013)Setny, Baron, Kekenes-Huskey, McCammon, and
  Dzubiella]{Setny:PNAS:2013}
Setny,~P.; Baron,~R.; Kekenes-Huskey,~P.~M.; McCammon,~J.~A.; Dzubiella,~J.
  \emph{Proc. Natl. Acad. Sci. U.S.A.} \textbf{2013}, \emph{110},
  1197--1202\relax
\mciteBstWouldAddEndPuncttrue
\mciteSetBstMidEndSepPunct{\mcitedefaultmidpunct}
{\mcitedefaultendpunct}{\mcitedefaultseppunct}\relax
\EndOfBibitem
\bibitem[Wei{\ss} \latin{et~al.}(2017)Wei{\ss}, Setny, and
  Dzubiella]{weiss2017principles}
Wei{\ss},~R.~G.; Setny,~P.; Dzubiella,~J. \emph{J. Chem. Theory Comput.}
  \textbf{2017}, \emph{13}, 3012--3019\relax
\mciteBstWouldAddEndPuncttrue
\mciteSetBstMidEndSepPunct{\mcitedefaultmidpunct}
{\mcitedefaultendpunct}{\mcitedefaultseppunct}\relax
\EndOfBibitem
\bibitem[Giacomello \latin{et~al.}(2015)Giacomello, Meloni, M{\"u}ller, and
  Casciola]{Giacomello:JCP:2015}
Giacomello,~A.; Meloni,~S.; M{\"u}ller,~M.; Casciola,~C.~M. \emph{J. Chem.
  Phys.} \textbf{2015}, \emph{142}, 104701\relax
\mciteBstWouldAddEndPuncttrue
\mciteSetBstMidEndSepPunct{\mcitedefaultmidpunct}
{\mcitedefaultendpunct}{\mcitedefaultseppunct}\relax
\EndOfBibitem
\bibitem[Menzl \latin{et~al.}(2016)Menzl, Gonzalez, Geiger, Caupin, Abascal,
  Valeriani, and Dellago]{menzl2016molecular}
Menzl,~G.; Gonzalez,~M.~A.; Geiger,~P.; Caupin,~F.; Abascal,~J.~L.;
  Valeriani,~C.; Dellago,~C. \emph{Proc. Natl. Acad. Sci. U.S.A.}
  \textbf{2016}, \emph{113}, 13582--13587\relax
\mciteBstWouldAddEndPuncttrue
\mciteSetBstMidEndSepPunct{\mcitedefaultmidpunct}
{\mcitedefaultendpunct}{\mcitedefaultseppunct}\relax
\EndOfBibitem
\bibitem[Zanjani \latin{et~al.}(2016)Zanjani, Jenkins, Crocker, and
  Sinno]{zanjani2016colloidal}
Zanjani,~M.~B.; Jenkins,~I.~C.; Crocker,~J.~C.; Sinno,~T. \emph{ACS Nano}
  \textbf{2016}, \emph{10}, 11280--11289\relax
\mciteBstWouldAddEndPuncttrue
\mciteSetBstMidEndSepPunct{\mcitedefaultmidpunct}
{\mcitedefaultendpunct}{\mcitedefaultseppunct}\relax
\EndOfBibitem
\bibitem[Kundu \latin{et~al.}(2011)Kundu, Sabhapandit, and
  Dhar]{kundu2011application}
Kundu,~A.; Sabhapandit,~S.; Dhar,~A. \emph{Phys. Rev. E} \textbf{2011},
  \emph{83}, 031119\relax
\mciteBstWouldAddEndPuncttrue
\mciteSetBstMidEndSepPunct{\mcitedefaultmidpunct}
{\mcitedefaultendpunct}{\mcitedefaultseppunct}\relax
\EndOfBibitem
\bibitem[Ray \latin{et~al.}(2017)Ray, Chan, and Limmer]{ray2017importance}
Ray,~U.; Chan,~G.~K.; Limmer,~D.~T. \emph{arXiv preprint} \textbf{2017},
  arXiv:1708.00459\relax
\mciteBstWouldAddEndPuncttrue
\mciteSetBstMidEndSepPunct{\mcitedefaultmidpunct}
{\mcitedefaultendpunct}{\mcitedefaultseppunct}\relax
\EndOfBibitem
\bibitem[Hassanali \latin{et~al.}(2013)Hassanali, Giberti, Cuny, K{\"u}hne, and
  Parrinello]{hassanali2013proton}
Hassanali,~A.; Giberti,~F.; Cuny,~J.; K{\"u}hne,~T.~D.; Parrinello,~M.
  \emph{Proc. Natl. Acad. Sci. U.S.A.} \textbf{2013}, \emph{110},
  13723--13728\relax
\mciteBstWouldAddEndPuncttrue
\mciteSetBstMidEndSepPunct{\mcitedefaultmidpunct}
{\mcitedefaultendpunct}{\mcitedefaultseppunct}\relax
\EndOfBibitem
\bibitem[Remsing \latin{et~al.}(2017)Remsing, Klein, and
  Sun]{remsing2017dependence}
Remsing,~R.~C.; Klein,~M.~L.; Sun,~J. \emph{Phys. Rev. B} \textbf{2017},
  \emph{96}, 024203\relax
\mciteBstWouldAddEndPuncttrue
\mciteSetBstMidEndSepPunct{\mcitedefaultmidpunct}
{\mcitedefaultendpunct}{\mcitedefaultseppunct}\relax
\EndOfBibitem
\bibitem[Sosso \latin{et~al.}(2016)Sosso, Caravati, Rotskoff, Vaikuntanathan,
  and Hassanali]{sosso2016role}
Sosso,~G.~C.; Caravati,~S.; Rotskoff,~G.; Vaikuntanathan,~S.; Hassanali,~A.
  \emph{J. Phys. Chem. A} \textbf{2016}, \emph{121}, 370--380\relax
\mciteBstWouldAddEndPuncttrue
\mciteSetBstMidEndSepPunct{\mcitedefaultmidpunct}
{\mcitedefaultendpunct}{\mcitedefaultseppunct}\relax
\EndOfBibitem
\bibitem[Xi \latin{et~al.}(2016)Xi, Remsing, and Patel]{Xi:JCTC:2016}
Xi,~E.; Remsing,~R.~C.; Patel,~A.~J. \emph{J. Chem. Theory Comput.}
  \textbf{2016}, \emph{12}, 706--713\relax
\mciteBstWouldAddEndPuncttrue
\mciteSetBstMidEndSepPunct{\mcitedefaultmidpunct}
{\mcitedefaultendpunct}{\mcitedefaultseppunct}\relax
\EndOfBibitem
\bibitem[Remsing \latin{et~al.}(2014)Remsing, Baer, Schenter, Mundy, and
  Weeks]{remsing2014role}
Remsing,~R.~C.; Baer,~M.~D.; Schenter,~G.~K.; Mundy,~C.~J.; Weeks,~J.~D.
  \emph{J. Phys. Chem. Lett.} \textbf{2014}, \emph{5}, 2767--2774\relax
\mciteBstWouldAddEndPuncttrue
\mciteSetBstMidEndSepPunct{\mcitedefaultmidpunct}
{\mcitedefaultendpunct}{\mcitedefaultseppunct}\relax
\EndOfBibitem
\bibitem[Patel \latin{et~al.}(2011)Patel, Varilly, Chandler, and
  Garde]{Patel:JSP:2011}
Patel,~A.~J.; Varilly,~P.; Chandler,~D.; Garde,~S. \emph{J. Stat. Phys.}
  \textbf{2011}, \emph{145}, 265 -- 275\relax
\mciteBstWouldAddEndPuncttrue
\mciteSetBstMidEndSepPunct{\mcitedefaultmidpunct}
{\mcitedefaultendpunct}{\mcitedefaultseppunct}\relax
\EndOfBibitem
\bibitem[Kumar \latin{et~al.}(1992)Kumar, Rosenberg, Bouzida, Swendsen, and
  Kollman]{Kumar:JCC:1992}
Kumar,~S.; Rosenberg,~J.~M.; Bouzida,~D.; Swendsen,~R.~H.; Kollman,~P.~A.
  \emph{J. Comp. Chem.} \textbf{1992}, \emph{13}, 1011--1021\relax
\mciteBstWouldAddEndPuncttrue
\mciteSetBstMidEndSepPunct{\mcitedefaultmidpunct}
{\mcitedefaultendpunct}{\mcitedefaultseppunct}\relax
\EndOfBibitem
\bibitem[Souaille and Roux(2001)Souaille, and Roux]{Roux:CPC:2001}
Souaille,~M.; Roux,~B. \emph{Computer Phys. Comm.} \textbf{2001}, \emph{135},
  40 -- 57\relax
\mciteBstWouldAddEndPuncttrue
\mciteSetBstMidEndSepPunct{\mcitedefaultmidpunct}
{\mcitedefaultendpunct}{\mcitedefaultseppunct}\relax
\EndOfBibitem
\bibitem[Shirts and Chodera(2008)Shirts, and Chodera]{MBAR}
Shirts,~M.~R.; Chodera,~J.~D. \emph{J. Chem. Phys.} \textbf{2008}, \emph{129},
  124105\relax
\mciteBstWouldAddEndPuncttrue
\mciteSetBstMidEndSepPunct{\mcitedefaultmidpunct}
{\mcitedefaultendpunct}{\mcitedefaultseppunct}\relax
\EndOfBibitem
\bibitem[Zhu and Hummer(2012)Zhu, and Hummer]{zhu2012convergence}
Zhu,~F.; Hummer,~G. \emph{J. Comput. Chem.} \textbf{2012}, \emph{33},
  453--465\relax
\mciteBstWouldAddEndPuncttrue
\mciteSetBstMidEndSepPunct{\mcitedefaultmidpunct}
{\mcitedefaultendpunct}{\mcitedefaultseppunct}\relax
\EndOfBibitem
\bibitem[Tan \latin{et~al.}(2012)Tan, Gallichio, Lapelosa, and Levy]{UWHAM}
Tan,~Z.; Gallichio,~E.; Lapelosa,~M.; Levy,~R.~M. \emph{J. Chem. Phys.}
  \textbf{2012}, \emph{136}, 144102\relax
\mciteBstWouldAddEndPuncttrue
\mciteSetBstMidEndSepPunct{\mcitedefaultmidpunct}
{\mcitedefaultendpunct}{\mcitedefaultseppunct}\relax
\EndOfBibitem
\bibitem[Stelzl \latin{et~al.}(2017)Stelzl, Kells, Rosta, and
  Hummer]{stelzl2017dynamic}
Stelzl,~L.~S.; Kells,~A.; Rosta,~E.; Hummer,~G. \emph{J. Chem. Theory Comput.}
  \textbf{2017}, \emph{13}, 6328--6342\relax
\mciteBstWouldAddEndPuncttrue
\mciteSetBstMidEndSepPunct{\mcitedefaultmidpunct}
{\mcitedefaultendpunct}{\mcitedefaultseppunct}\relax
\EndOfBibitem
\bibitem[Remsing and Patel(2015)Remsing, and Patel]{Remsing:JCP:2015}
Remsing,~R.~C.; Patel,~A.~J. \emph{J. Chem. Phys.} \textbf{2015}, \emph{142},
  024502\relax
\mciteBstWouldAddEndPuncttrue
\mciteSetBstMidEndSepPunct{\mcitedefaultmidpunct}
{\mcitedefaultendpunct}{\mcitedefaultseppunct}\relax
\EndOfBibitem
\bibitem[Patel and Garde(2014)Patel, and Garde]{Patel:JPCB:2014}
Patel,~A.~J.; Garde,~S. \emph{J. Phys. Chem. B} \textbf{2014}, \emph{118},
  1564--1573\relax
\mciteBstWouldAddEndPuncttrue
\mciteSetBstMidEndSepPunct{\mcitedefaultmidpunct}
{\mcitedefaultendpunct}{\mcitedefaultseppunct}\relax
\EndOfBibitem
\bibitem[Pohorille \latin{et~al.}(2010)Pohorille, Jarzynski, and
  Chipot]{good_practices}
Pohorille,~A.; Jarzynski,~C.; Chipot,~C. \emph{J. Phys. Chem. B} \textbf{2010},
  \emph{114}, 10235--10253\relax
\mciteBstWouldAddEndPuncttrue
\mciteSetBstMidEndSepPunct{\mcitedefaultmidpunct}
{\mcitedefaultendpunct}{\mcitedefaultseppunct}\relax
\EndOfBibitem
\bibitem[K{\"a}stner and Thiel(2005)K{\"a}stner, and Thiel]{Kastner:JCP:2005}
K{\"a}stner,~J.; Thiel,~W. \emph{J. Chem. Phys.} \textbf{2005}, \emph{123},
  144104\relax
\mciteBstWouldAddEndPuncttrue
\mciteSetBstMidEndSepPunct{\mcitedefaultmidpunct}
{\mcitedefaultendpunct}{\mcitedefaultseppunct}\relax
\EndOfBibitem
\bibitem[Bennett(1976)]{BAR}
Bennett,~C.~H. \emph{J. Comput. Phys.} \textbf{1976}, \emph{22}, 245--268\relax
\mciteBstWouldAddEndPuncttrue
\mciteSetBstMidEndSepPunct{\mcitedefaultmidpunct}
{\mcitedefaultendpunct}{\mcitedefaultseppunct}\relax
\EndOfBibitem
\bibitem[Hummer \latin{et~al.}(1996)Hummer, Garde, Garcia, Pohorille, and
  Pratt]{Hummer:PNAS:1996}
Hummer,~G.; Garde,~S.; Garcia,~A.~E.; Pohorille,~A.; Pratt,~L.~R. \emph{Proc.
  Natl. Acad. Sci. U.S.A.} \textbf{1996}, \emph{93}, 8951--8955\relax
\mciteBstWouldAddEndPuncttrue
\mciteSetBstMidEndSepPunct{\mcitedefaultmidpunct}
{\mcitedefaultendpunct}{\mcitedefaultseppunct}\relax
\EndOfBibitem
\bibitem[Garde \latin{et~al.}(1996)Garde, Hummer, Garcia, Paulaitis, and
  Pratt]{garde96}
Garde,~S.; Hummer,~G.; Garcia,~A.~E.; Paulaitis,~M.~E.; Pratt,~L.~R.
  \emph{Phys. Rev. Lett.} \textbf{1996}, \emph{77}, 4966--4968\relax
\mciteBstWouldAddEndPuncttrue
\mciteSetBstMidEndSepPunct{\mcitedefaultmidpunct}
{\mcitedefaultendpunct}{\mcitedefaultseppunct}\relax
\EndOfBibitem
\bibitem[Hummer \latin{et~al.}(1998)Hummer, Garde, Garcia, Paulaitis, and
  Pratt]{hummer98pnas}
Hummer,~G.; Garde,~S.; Garcia,~A.; Paulaitis,~M.; Pratt,~L. \emph{Proc. Natl.
  Acad. Sci. U.S.A.} \textbf{1998}, \emph{95}, 1552--1555\relax
\mciteBstWouldAddEndPuncttrue
\mciteSetBstMidEndSepPunct{\mcitedefaultmidpunct}
{\mcitedefaultendpunct}{\mcitedefaultseppunct}\relax
\EndOfBibitem
\bibitem[Sarupria and Garde(2009)Sarupria, and Garde]{garde09prl}
Sarupria,~S.; Garde,~S. \emph{Phys. Rev. Lett.} \textbf{2009}, \emph{103},
  037803\relax
\mciteBstWouldAddEndPuncttrue
\mciteSetBstMidEndSepPunct{\mcitedefaultmidpunct}
{\mcitedefaultendpunct}{\mcitedefaultseppunct}\relax
\EndOfBibitem
\bibitem[Garde and Patel(2011)Garde, and Patel]{Garde:PNAS:2011}
Garde,~S.; Patel,~A.~J. \emph{Proc. Natl. Acad. Sci. U.S.A.} \textbf{2011},
  \emph{108}, 16491--16492\relax
\mciteBstWouldAddEndPuncttrue
\mciteSetBstMidEndSepPunct{\mcitedefaultmidpunct}
{\mcitedefaultendpunct}{\mcitedefaultseppunct}\relax
\EndOfBibitem
\bibitem[Sanyal and Shell(2016)Sanyal, and Shell]{sanyal2016coarse}
Sanyal,~T.; Shell,~M.~S. \emph{The Journal of chemical physics} \textbf{2016},
  \emph{145}, 034109\relax
\mciteBstWouldAddEndPuncttrue
\mciteSetBstMidEndSepPunct{\mcitedefaultmidpunct}
{\mcitedefaultendpunct}{\mcitedefaultseppunct}\relax
\EndOfBibitem
\bibitem[Zheng \latin{et~al.}(1991)Zheng, Durben, Wolf, and
  Angell]{zheng1991liquids}
Zheng,~Q.; Durben,~D.; Wolf,~G.; Angell,~C. \emph{Science} \textbf{1991},
  \emph{254}, 829--832\relax
\mciteBstWouldAddEndPuncttrue
\mciteSetBstMidEndSepPunct{\mcitedefaultmidpunct}
{\mcitedefaultendpunct}{\mcitedefaultseppunct}\relax
\EndOfBibitem
\bibitem[Pallares \latin{et~al.}(2014)Pallares, Azouzi, Gonz{\'a}lez, Aragones,
  Abascal, Valeriani, and Caupin]{pallares2014anomalies}
Pallares,~G.; Azouzi,~M. E.~M.; Gonz{\'a}lez,~M.~A.; Aragones,~J.~L.;
  Abascal,~J.~L.; Valeriani,~C.; Caupin,~F. \emph{Proc. Natl. Acad. Sci.
  U.S.A.} \textbf{2014}, \emph{111}, 7936--7941\relax
\mciteBstWouldAddEndPuncttrue
\mciteSetBstMidEndSepPunct{\mcitedefaultmidpunct}
{\mcitedefaultendpunct}{\mcitedefaultseppunct}\relax
\EndOfBibitem
\bibitem[Biddle \latin{et~al.}(2017)Biddle, Singh, Sparano, Ricci,
  Gonz{\'a}lez, Valeriani, Abascal, Debenedetti, Anisimov, and
  Caupin]{biddle2017two}
Biddle,~J.~W.; Singh,~R.~S.; Sparano,~E.~M.; Ricci,~F.; Gonz{\'a}lez,~M.~A.;
  Valeriani,~C.; Abascal,~J.~L.; Debenedetti,~P.~G.; Anisimov,~M.~A.;
  Caupin,~F. \emph{J. Chem. Phys.} \textbf{2017}, \emph{146}, 034502\relax
\mciteBstWouldAddEndPuncttrue
\mciteSetBstMidEndSepPunct{\mcitedefaultmidpunct}
{\mcitedefaultendpunct}{\mcitedefaultseppunct}\relax
\EndOfBibitem
\bibitem[Cisneros \latin{et~al.}(2016)Cisneros, Wikfeldt, Ojam{\"a}e, Lu, Xu,
  Torabifard, Bart{\'o}k, Cs{\'a}nyi, Molinero, and
  Paesani]{cisneros2016modeling}
Cisneros,~G.~A.; Wikfeldt,~K.~T.; Ojam{\"a}e,~L.; Lu,~J.; Xu,~Y.;
  Torabifard,~H.; Bart{\'o}k,~A.~P.; Cs{\'a}nyi,~G.; Molinero,~V.; Paesani,~F.
  \emph{Chem. Rev.} \textbf{2016}, \emph{116}, 7501--7528\relax
\mciteBstWouldAddEndPuncttrue
\mciteSetBstMidEndSepPunct{\mcitedefaultmidpunct}
{\mcitedefaultendpunct}{\mcitedefaultseppunct}\relax
\EndOfBibitem
\bibitem[Morawietz \latin{et~al.}(2016)Morawietz, Singraber, Dellago, and
  Behler]{morawietz2016van}
Morawietz,~T.; Singraber,~A.; Dellago,~C.; Behler,~J. \emph{Proc. Natl. Acad.
  Sci. U.S.A.} \textbf{2016}, \emph{113}, 8368--8373\relax
\mciteBstWouldAddEndPuncttrue
\mciteSetBstMidEndSepPunct{\mcitedefaultmidpunct}
{\mcitedefaultendpunct}{\mcitedefaultseppunct}\relax
\EndOfBibitem
\bibitem[Awasthi \latin{et~al.}(2016)Awasthi, Kapil, and
  Nair]{awasthi2016sampling}
Awasthi,~S.; Kapil,~V.; Nair,~N.~N. \emph{J. Comp. Chem.} \textbf{2016},
  \emph{37}, 1413--1424\relax
\mciteBstWouldAddEndPuncttrue
\mciteSetBstMidEndSepPunct{\mcitedefaultmidpunct}
{\mcitedefaultendpunct}{\mcitedefaultseppunct}\relax
\EndOfBibitem
\bibitem[Awasthi and Nair(2017)Awasthi, and Nair]{awasthi2017exploring}
Awasthi,~S.; Nair,~N.~N. \emph{J. Chem. Phys.} \textbf{2017}, \emph{146},
  094108\relax
\mciteBstWouldAddEndPuncttrue
\mciteSetBstMidEndSepPunct{\mcitedefaultmidpunct}
{\mcitedefaultendpunct}{\mcitedefaultseppunct}\relax
\EndOfBibitem
\bibitem[Pfaendtner and Bonomi(2015)Pfaendtner, and
  Bonomi]{pfaendtner2015efficient}
Pfaendtner,~J.; Bonomi,~M. \emph{J. Chem. Theory Comput.} \textbf{2015},
  \emph{11}, 5062--5067\relax
\mciteBstWouldAddEndPuncttrue
\mciteSetBstMidEndSepPunct{\mcitedefaultmidpunct}
{\mcitedefaultendpunct}{\mcitedefaultseppunct}\relax
\EndOfBibitem
\bibitem[Hess \latin{et~al.}(2008)Hess, Kutzner, van~der Spoel, and
  Lindahl]{gmx4ref}
Hess,~B.; Kutzner,~C.; van~der Spoel,~D.; Lindahl,~E. \emph{J. Chem. Theory
  Comput.} \textbf{2008}, 435 -- 447\relax
\mciteBstWouldAddEndPuncttrue
\mciteSetBstMidEndSepPunct{\mcitedefaultmidpunct}
{\mcitedefaultendpunct}{\mcitedefaultseppunct}\relax
\EndOfBibitem
\bibitem[Frenkel and Smit(2002)Frenkel, and Smit]{Frenkel_Smit}
Frenkel,~D.; Smit,~B. \emph{Understanding Molecular Simulations: From
  Algorithms to Applications}, 2nd ed.; Academic Press, New York, 2002\relax
\mciteBstWouldAddEndPuncttrue
\mciteSetBstMidEndSepPunct{\mcitedefaultmidpunct}
{\mcitedefaultendpunct}{\mcitedefaultseppunct}\relax
\EndOfBibitem
\bibitem[Berendsen \latin{et~al.}(1987)Berendsen, Grigera, and Straatsma]{SPCE}
Berendsen,~H. J.~C.; Grigera,~J.~R.; Straatsma,~T.~P. \emph{J. Phys. Chem.}
  \textbf{1987}, \emph{91}, 6269--6271\relax
\mciteBstWouldAddEndPuncttrue
\mciteSetBstMidEndSepPunct{\mcitedefaultmidpunct}
{\mcitedefaultendpunct}{\mcitedefaultseppunct}\relax
\EndOfBibitem
\bibitem[Ryckaert \latin{et~al.}(1977)Ryckaert, Ciccotti, and Berendsen]{SHAKE}
Ryckaert,~J.-P.; Ciccotti,~G.; Berendsen,~H. J.~C. \emph{J. Comp. Phys.}
  \textbf{1977}, \emph{23}, 327 -- 341\relax
\mciteBstWouldAddEndPuncttrue
\mciteSetBstMidEndSepPunct{\mcitedefaultmidpunct}
{\mcitedefaultendpunct}{\mcitedefaultseppunct}\relax
\EndOfBibitem
\bibitem[Essmann \latin{et~al.}(1995)Essmann, Perera, Berkowitz, Darden, Lee,
  and Pedersen]{PME}
Essmann,~U.; Perera,~L.; Berkowitz,~M.~L.; Darden,~T.; Lee,~H.; Pedersen,~L.~G.
  \emph{J. Chem. Phys.} \textbf{1995}, \emph{103}, 8577--8593\relax
\mciteBstWouldAddEndPuncttrue
\mciteSetBstMidEndSepPunct{\mcitedefaultmidpunct}
{\mcitedefaultendpunct}{\mcitedefaultseppunct}\relax
\EndOfBibitem
\bibitem[Bussi \latin{et~al.}(2007)Bussi, Donadio, and Parrinello]{V-Rescale}
Bussi,~G.; Donadio,~D.; Parrinello,~M. \emph{J. Chem. Phys.} \textbf{2007},
  \emph{126}, 014101\relax
\mciteBstWouldAddEndPuncttrue
\mciteSetBstMidEndSepPunct{\mcitedefaultmidpunct}
{\mcitedefaultendpunct}{\mcitedefaultseppunct}\relax
\EndOfBibitem
\bibitem[Parrinello and Rahman(1981)Parrinello, and Rahman]{Parrinello-Rahman}
Parrinello,~M.; Rahman,~A. \emph{J. Applied Phys.} \textbf{1981}, \emph{52},
  7182--7190\relax
\mciteBstWouldAddEndPuncttrue
\mciteSetBstMidEndSepPunct{\mcitedefaultmidpunct}
{\mcitedefaultendpunct}{\mcitedefaultseppunct}\relax
\EndOfBibitem
\end{mcitethebibliography}



\setlength{\baselineskip}{18pt}

\clearpage

\huge \textbf{Supporting Information} \normalsize

SI has been appended below.

\normalsize

\section{Relating Free Energetics in the Biased and Unbiased Ensembles}

%
Here we include a derivation of Equation 1 in the main text, which often serves as the starting point for enhanced sampling methods such as umbrella sampling.
The probability, $P_v(\tilde N)$, of observing $\tilde N$ coarse-grained waters in the unbiased ensemble with the generalized Hamiltonian, $\mathcal{H}_0$, is given by:
\begin{equation}
P_v(\tilde N) = \langle \delta(\tilde N_v- \tilde N) \rangle _0 = \frac{1}{Q_0} \int d\Rbar~\delta(\tilde N_v- \tilde N) e^{-\beta \mathcal{H}_0(\Rbar)}
\label{eq:pvn-def}
\end{equation}
where $Q_0\equiv\int d\Rbar~e^{-\beta \mathcal{H}_0(\Rbar)}$ is the partition function of the unbiased ensemble, 
and $\Rbar$ represents the positions of all atoms in the system.
Upon addition of a biasing potential, the statistics of the biased ensemble are dictated by the Hamiltonian, $\mathcal{H}_{\bar \lambda}(\Rbar) = \mathcal{H}_0(\Rbar) + U_{\bar \lambda} (\tilde N_v(\Rbar))$.
In particular, the biased distribution,
\begin{equation}
P_v^{\bar \lambda}(\tilde N) = \langle \delta(\tilde N_v- \tilde N) \rangle _{\bar \lambda} = \frac{1}{Q_{\bar \lambda}} \int d\Rbar~\delta(\tilde N_v- \tilde N) e^{-\beta \mathcal{H}_{\bar \lambda}(\Rbar)},
\label{eq:pvnb-def}
\end{equation}
where $Q_{\bar \lambda}\equiv\int d\Rbar~e^{-\beta \mathcal{H}_{\bar \lambda}(\Rbar)}$ is the partition function of the biased ensemble.
To relate $P_v(\tilde N)$ and $P_v^{\bar \lambda}(\tilde N)$, we can rewrite Equation~\ref{eq:pvn-def} as:
\begin{equation}
P_v(\tilde N) = \left(\frac{Q_{\bar \lambda}}{Q_0}\right)\frac{1}{Q_{\bar\lambda}} \int d\Rbar~\delta(\tilde N_v- \tilde N) e^{-\beta \mathcal{H}_{\bar \lambda}} e^{\beta U_{\bar \lambda}({\tilde N}_v)}
\label{eq:pvn-rwt}
\end{equation}
Recognizing that the delta function allows us to rewrite $e^{\beta U_{\bar \lambda}({\tilde N}_v)}$ as $e^{\beta U_{\bar \lambda}({\tilde N})}$, and pull it outside the integral (because it no longer depends on $\Rbar$), we get
\begin{equation}
P_v(\tilde N) = \left(\frac{Q_{\bar \lambda}}{Q_0}\right) e^{\beta U_{\bar \lambda}({\tilde N})}   \frac{1}{Q_{\bar\lambda}} \int d\Rbar~\delta(\tilde N_v- \tilde N) e^{-\beta \mathcal{H}_{\bar \lambda}} 
\label{eq:pvn-rwt2}
\end{equation}
Using Equation~\ref{eq:pvnb-def} and defining the free energy difference between the biased and the unbiased ensembles, $F_{\bar \lambda} \equiv (-1/\beta)\ln(Q_{\bar \lambda}/Q_0)$, we then get:
\begin{equation}
P_v(\tilde N) = e^{-\beta F_{\bar \lambda}} e^{\beta U_{\bar \lambda}(\tilde N)} P_v^{\bar \lambda}(\tilde N)
\label{eq:pvn-final}
\end{equation}
Finally, taking the logarithm of both sides and multiplying by $-(1/\beta)$, we get the desired result:
\begin{equation}
F_v(\tilde N) = F_v^{\bar \lambda} (\tilde N) - U_{\bar \lambda} (\tilde N) + F_{\bar \lambda}
\label{eq:pvn-ln}
\end{equation}

%
\section{Comparing Sparse Sampling to Umbrella Integration}
%
Here, we compare sparse sampling with the umbrella integration (UI) method of Kastner and Thiel~\cite{Kastner:JCP:2005, Kastner:JCP:2006}.
Although both methods employ biased simulations in conjunction with thermodynamic integration to obtain free energy profiles,
there are important differences both in the spirit of the two methods, and in their respective implementations. 
In lieu of estimating the offset, $F_{\kappa,N^*}$, UI seeks to eliminate it from the equation by considering the derivative of Equation 4 of the main text.
Thus, the central equation for UI becomes:
$$ \frac{\partial F_v}{\partial \tilde N} =  \frac{\partial F_v^{\kappa,N^*}}{\partial \tilde N} - \frac{\partial U_{\kappa,N^*}}{\partial \tilde N} = \frac{\partial F_v^{\kappa,N^*}}{\partial \tilde N} - \kappa ( \tilde N - N^*). $$
In UI, the biased free energetics, $F_v^{\kappa,N^*}(\tilde N)$ are further assumed to be parabolic, 
so that they can be estimated using the mean and the variance of the biased distributions.
Estimates of $\partial F_v / \partial \tilde N$ obtained from different biased distributions are then combined using a sensible weighted averaging scheme, and integrated numerically to obtain estimates of $F_v(\tilde N)$.
Formulated in this way, UI was proposed to circumvent many of the early challenges associated with WHAM, 
such as the choice of bin width or the estimation of errors in $F_{\kappa,N^*}$ (these issues have since been addressed 
by more recent reformulations of WHAM, such as MBAR~\cite{MBAR} or UWHAM~\cite{UWHAM}.)
By using only the biased means and variances, 
whose convergences could be readily ascertained, 
UI provided a useful alternative for analyzing umbrella sampling simulations.

In contrast with UI, sparse sampling seeks to estimate $F_v(\tilde N)$ using a small number of short simulations 
that are sparsely distributed across the order parameter space. 
In this way, sparse sampling seeks to circumvent the overlap requirement intrinsic to umbrella sampling, 
and provide a computationally efficient alternative for estimating free energy landscapes.
In sparse sampling, $F_{\kappa,N^*}$ is explicitly estimated by integrating its derivative, $dF_{\kappa,N^*}/dN^*$ with respect to $N^*$ 
(in contrast, $\partial F_v / \partial \tilde N$ is integrated with respect to $\tilde N$ in UI).
Thus, we rely on a knowledge of the functional form of $dF_{\kappa,N^*}/dN^*$ on $N^*$, which is 
determined approximately at first, and is refined further using subsequent simulations.
Finally, we note that as $\kappa\to\infty$, the two method converge; 
however, as discussed in the main text, errors in $dF_{\kappa,N^*}/dN^*$ increase with increasing $\kappa$,
making it expedient to choose a small $\kappa$-value that is nevertheless
large enough to result in monostable biased free energy profiles.
To facilitate such an optimal choice of $\kappa$, sparse sampling makes use of estimates of $\partial F_v / \partial \tilde N$ at $\tilde N = \langle N_v \rangle_{\kappa,N^*}$.


\end{document}